%% file: JADE-EPJH20.tex
\documentclass{sn-jnl}
\usepackage{xcolor}
\usepackage[normalem]{ulem}
\def\oq{\char'134}

\def\z0{\rm Z^0}
\newcommand{\as}{\alpha_{\rm s}}
\newcommand{\asq}{\alpha_{\rm s}(q)}

\newcommand{\oaa}{{\cal O}(\as^2)}
\newcommand{\oaaa}{{\cal O}(\as^3)}
\newcommand{\epem}{\rm e^+\rm e^-}
\newcommand{\amz}{\as(M_{\rm Z^0})}

\begin{document}
\title[The JADE Experiment]{The JADE Experiment at the PETRA $e^+e^-$ collider -
history, achievements and revival}      

\author*[1]{\fnm{S.} \sur{Bethke}}\email{bethke@mpp.mpg.de}

\author*[2]{\fnm{A.} \sur{Wagner}} \email{albrecht.wagner@desy.de}

\affil[1]{\orgname{Max-Planck-Institute of Physics}, \orgaddress{\street{F\"ohringer Ring 6}, \city{Munich},
\postcode{80805}, \state{Germany}}}

\affil[2]{\orgname{DESY}, \orgaddress{\street{Notkestr. 85}, \city{Hamburg},
\postcode{22607}, \state{Germany}}}

\abstract{
The JADE experiment was one of five large detector systems taking data
at the electron-positron collider PETRA, from 1979 to 1986,
at $e^+e^-$ annihilation centre-of-mass energies from 12 to 46.7 GeV.
The forming of the JADE collaboration, the construction
of the apparatus, the most prominent physics highlights, 
and the post-mortem resurrection and preservation of JADE's data and software 
are reviewed.
} 

\maketitle

\tableofcontents

\input{Sec-01} 
\input{Sec-02} 
\input{Sec-03} 
\input{Sec-04} 
\input{Sec-05} 
%

\input{references}
\end{document}

%% file: Sec-01.tex
\section{Introduction}
\label{intro}
JADE was a particle detector system and experiment at the electron-positron 
collider PETRA at the DESY laboratory in Hamburg, Germany.
It was designed, constructed and operated by an international collaboration of institutes from 
{\bf JA}pan, {\bf D}eutschland and {\bf E}ngland \cite{jade-all}.
From 1979 to 1986, the experiment recorded data of electron-positron annihilations at 
centre-of-mass energies between 12 and 46.7 GeV. 

The detector comprised novel technologies like accurate tracking of charged particles in the central jet chamber with fast multi-hit electronics, measurement and identification of photons, electrons and muons over wide regions of solid angle, and maximal hermiticity and symmetry of the detector systems. 
It was designed to identify and precisely measure the dynamics of
electrons, muons, photons, hadrons and hadron jets
in $\epem$ annihilation final states, and thus to study and test the underlying 
theoretical framework of the Standard Model (SM), the unified theory of electromagnetic and
weak interaction, to study hadronic final states and the dynamics of the strong interaction,
and to search for new particles.

Scientific highlight results of JADE
were the (co-)discovery of the gluon, the development of jet-physics and first 
evidence for the asymptotic freedom of quarks and gluons,
first evidence for phenomena through electro-weak interference, 
measurements of the photon structure function 
and searches for New Physics like Super-Symmetry, free quarks 
and the - at that time still undiscovered - top-quark.

More than 10 years after close-down of the experiment, the JADE software and data were 
revived and rescued to modern storage systems and computing platforms, 
showcasing the
need for data preservation and re-use in particle physics and beyond \cite{jade-revival}.
These post-mortem efforts provided the means for a second series of PhD-theses and 
scientific publications, based on novel analysis methods and improved theoretical
calculations and models that were not available during the life-time of the experiment.

In this article, the history and the development of the JADE experiment, the revival 
and preservation of its data and software, and some of its major scientific achievements 
are being reviewed.

\section{Historical context}
\label{sec:1}
About 50 years ago, particle physics was boosted by a sequence of groundbreaking 
experimental discoveries and theoretical developments 
{that had major impact on preparing JADE:
  
In the late 1960's and early 1970's, the discovery of the proton's substructure in experiments
scattering highly energetic electrons off nucleons, at the Stanford 
Linear Accelerator Center (SLAC) in California, confirmed the predictions of the parton model 
\cite{gell-mann,zweig,feynman-partons,bjorken-partons};
see e.g \cite{friedman} for a review 
summarising the status of evidence in 1972. 
These developments established quarks as the fundamental
constituents of hadronic matter.
They were honoured by Nobel prices to M. Gell-Mann in 1969, for the proposal of
the quark model, and to J. Friedman, H. Kendall and R. Taylor in 1990,
for their pioneering investigations of deep inelastic electron scattering and the
establishment of the quark-parton model.
Three different kinds (\oq flavours") of quarks
- the up (u), the down (d) and the strange (s) quark - were proposed to compose 
the known  \oq Zoo" of hadrons at that time.
 
In 1973, the discovery of weak neutral currents in neutrino scattering experiments at CERN in Geneva / 
Switzerland provided experimental evidence 
\cite{nc1,neutralcurrents}
for the validity of the Standard Model of electro-weak interactions.
This model was developed since the 1960s 
\cite{glashow,weinberg,salam}, 
and its success was honoured by the Nobel price for
Sheldon Glashow, Abdus Salam and Steven Weinberg in 1979 \cite{nobel79}.

In November 1974, a fourth species of quarks, 
the charm-quark, was discovered  
by the E598 experiment at Brookhaven National Lab (BNL) in New York \cite{jpsi1}, 
and the SLAC-LBL magnetic detector at the electron-positron accelerator SPEAR
at the Stanford Linear Accelerator Centre (SLAC) in California \cite{jpsi2}.
Called the \oq November Revolution", this discovery triggered a series of new developments 
in the field of particle physics, and earned the Nobel price for Burton Richter (SLAC) 
and Samuel Ting (MIT) in 1976 \cite{nobel1976}.

The symmetry of two doublets of quarks and leptons 
as basic constituents of hadronic matter did not last very long: 
First evidence for the existence of a third, sequential lepton, the $\tau$-lepton, was reported 
in 1975 \cite{tau}, from data of the SLAC-LBL magnetic detector at SPEAR,
for which Martin L. Perl received the Nobel Prize in 1995 \cite{nobel1995}.
In 1977, a narrow resonance observed at Fermilab close to Chicago revealed the existence of 
a fifth species of quarks, called the b-, beauty- or bottom-quark
\cite{b-quark}.
This immediately opened the quest for the existence of the b-quark's doublet partner, the top-quark,
and boosted discussions and developments of plans for future particle accelerators and 
experiments.

The theory of the strong interaction between quarks, Quantum-Chromodynamics
(QCD), started to emerge in the 1960s, and was completed in the early
1970s by introducing colour octet gluons as carriers of the strong force \cite{nambu,gluons}, 
and by the concepts of asymptotic freedom 
\textcolor{blue}{\cite{gross,politzer}} 
and of the confinement of quarks, 
at high and at low energy scales, respectively. 
Based on QCD, the formation of observable 
hadron jets as footprints of quarks and hard gluon emission was predicted and quantified 
\cite{exp-gluons,jets}, paving the way for experimental verifications and further possible discoveries.

During this 
intense time of discoveries and the development
of modern particle physics,
many new projects for particle colliders were discussed and planned in Europe,
the United States, in Japan and in Russia. 
At DESY, the German laboratory for
particle physics at Hamburg, the $\epem$ collider DORIS had 
been commissioned in 1974, 
just after the \oq November Revolution", contributing with the observation 
of higher mass excitations of the newly discovered $J/\Psi$ charm-quark-antiquark states. 

In 1973, during a meeting of the European Committee for Future Accelerators, ECFA, 
about long-term plans for new accelerators in Europe it emerged that DESY, 
the Rutherford Appleton Laboratory
(then called RHEL) in the UK, and the INFN laboratory in Frascati/Italy considered construction 
of a large $\epem$ storage ring. 
There was consensus that only one such collider should be built in Europe. 
In 1975 the Administrative Council of DESY concluded that the plan of an electron-positron 
storage ring for energies around 20 GeV was scientifically very well founded, 
and that DESY offered the best conditions in Europe for the realisation of such a project, 
since the existing experimental facilities would be fully used. 
DESY was asked by the German Government
to continue its efforts to secure international scientific and financial support
and to prepare for the international use of the project.

Intense negotiations of the DESY Director Herwig Schopper with the German government 
led to the approval of the 
Positron-Electron-Tandem-Ring-Anlage at DESY, PETRA \cite{petra},
already in the fall of 1975.
The foundation for experiments at the PETRA $\epem$ collider was laid at the 
Discussion Meeting on PETRA Experiments, 
held at Frascati/Italy from March 1 to 5, 1976 \cite{frascati}.
At this meeting, presentations about the physics goals, technical requirements 
and proposals for experimental setups were given and discussed,
and first steps to form collaborations for  
experiments at PETRA were taken.

Under the leadership of Gustaf-Adolph Voss, the Director of Accelerators, 
who already had led the design efforts, construction
of PETRA began in 1976, and was completed in 1978. 
First stored beams were established in July 1978, six months ahead of the initial planning. 
The initial experiments operating at the 4 collision regions of PETRA 
were the JADE \cite{jade-total-x}, Mark-J \cite{markj}, PLUTO \cite{pluto} and TASSO \cite{tasso} detectors. 
PLUTO had already operated \cite{pluto-doris} at the precursor storage ring DORIS, 
and was replaced at PETRA in March 1980 by the newly built CELLO detector \cite{cello}.

The timeline of PETRA as $\epem$ storage ring and collider
is summarised in Table~\ref{tab:1}.

\begin{table}[htb]
\centering
\caption{Timeline of the $e^+e^-$ storage ring PETRA}
\label{tab:1}      
\begin{tabular}{rll}
\hline\noalign{\smallskip}
Date & & Event  \\
\noalign{\smallskip}\hline\noalign{\smallskip}
1974 & September: & proposal for an $\epem$ storage ring of 2304 m circumference, \\
 & &  with beam energies $E_{b}$ up to 19 GeV \\
 & & and luminosity up to $1.2 \times 10^{32} $\ cm$^{-2}$\ s$^{-1}$ \\
1975 & Fall :& approval by federal government \\
1976 & March 1-5: & discussion meeting on PETRA experiments, Frascati (Italy)\\
  & Spring:  & start of PETRA construction; formation of PETRA Research \\
  &  & Committee (PRC); call for experimental proposals \\
  & August: & 12 experimental proposals received \\
  & October: & 5 proposals recommended for approval (CELLO, JADE,  \\
  & & Mark-J, PLUTO, TASSO) \\
1977 & February: & PETRA magnet series production started \\
1978 & July: & early completion of PETRA, start of operation \\
  & September: & first $\epem$ collisions \\
  & October: & Mark-J, PLUTO, TASSO detectors moved into beam \\
1979 & January: & first experimental results; 
  JADE detector moved into beam \\
  & April 26: & official inauguration of PETRA with presence of the \\ 
  & & President of the Federal Republic of Germany, Walter Scheel \\
  & November: & $E_b$ up to 19 GeV \\ 
1983 & & Energy upgrade: $E_b$ up to 22 GeV \\
1984 & March: & maximum $E_b$ of 23.390 GeV \\
1986 & November 3: & final shutdown of PETRA and experiments \\
\noalign{\smallskip}\hline
\end{tabular}
\end{table}

After its shutdown as particle collider in November 1986, PETRA came back to life in 1990 as PETRA-II, 
where it served as pre-accelerator for the electron-proton collider HERA until 2007.
From 1995 PETRA-II also operated as synchrotron radiation source, and 
from 2007 to 2009 was converted 
into a highly brillant X-ray source, PETRA-III.
 
With PETRA and four large experimental setups starting to operate in 1979,
DESY became a nationally funded, yet internationally used research laboratory 
in the field of particle physics.
Fig.~\ref{fig:petra} schematically shows the PETRA collider, with its 4 experimental halls and experiment sites.

\begin{figure}[htb]
  \includegraphics[width=0.8\textwidth]{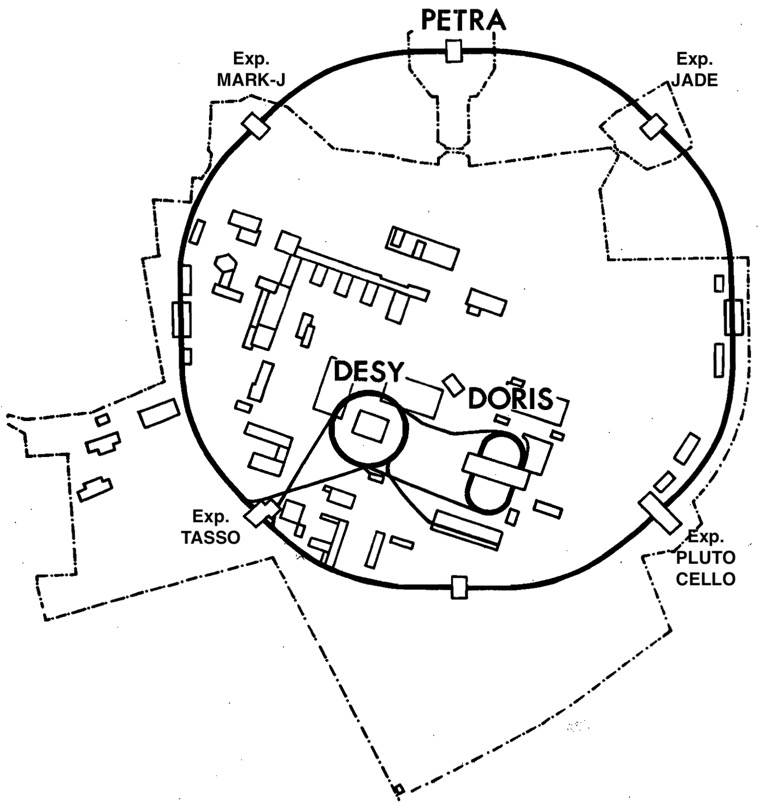}
\caption{PETRA and its 4 experimental halls on the DESY site (dash-dotted boundary).}
\label{fig:petra}       
\end{figure}

%% file: Sec-02.tex
\section{The JADE detector and collaboration - planning, set-up and operation}
\label{sec:2}
\subsection{Planning}
\label{sec:planning}

The initial founders of the JADE collaboration were Joachim Heintze, Professor at the
Physikalisches Institut of Heidelberg University, Masatoshi Koshiba, Professor at Tokyo University, and Shuji Orito, scientist at the Max-Planck-Institute for Physics at Munich and later head of the newly setup Laboratory of International Collaboration on Elementary Particle Physics (LICEPP) at Tokyo University.

At Frascati, Heintze, at that time spokesperson of the DESY-Heidelberg NaI Lead Glass detector operating at the
DORIS storage ring at DESY, presented a talk on \oq Ideas how to use the DESY-Heidelberg Equipment at PETRA", and pointed out the importance to measure as many parameters 
as possible of all particles emerging from the $\epem$ annihilation processes.
Orito, at that time member of the DASP experiment at DORIS, presented \oq A 4$\pi$ Detector with good Electron and Muon Identification", and also stressed the importance of obtaining the  most 
complete and precise information on the kinematics of PETRA collision events.
Furthermore, Robin Marshall from Daresbury Lab (UK) reported ideas on \oq A Magnetic Detector with Lepton and Photon Identification", and Wulfrin Bartel from DESY and the University of Hamburg, like Heintze member of the DESY-Heidelberg experiment, presented plans for an \oq Electron-Hadron Calorimeter for PETRA".

As these proposals were based on the same basic principles it was natural that
Heintze, Orito, Marshall, Bartel and colleagues joined forces and proposed
a new magnetic detector to be operated at PETRA, 
with maximum sensitivity for measuring the parameters of leptons and hadrons 
over (close to) the full solid angle.
Heintze and Orito also involved
Rolf Felst, senior scientist at DESY and head of the DESY F22 group, and member of DASP as well.

The initial plan included a high resolution detector for charged particles within a magnetic
coil of 30 cm radius - basically following the parameters of the DESY-Heidelberg experiment. 
Rolf Felst and Dieter Cords suggested a coil radius of at least 1 m.
Joachim Heintze and his group, being in charge for construction of the tracking chamber,
took up this proposal - which later proved to be an essential decision for the scientific success
of the detector.

The groups of Heidelberg, DESY and Tokyo were joined by groups 
from Daresbury Lab, from University of Lancaster and from the University of Manchester,
with Paul Murphy of Manchester University being the leading and most influential person
of the English groups.
Together, they set up the official \oq JADE - Proposal for a Compact Magnetic Detector at PETRA"
\cite{jade-proposal}, which 
was submitted to the DESY management by August 1976.
The name of the collaboration and experiment was derived from the country names of the
founding institutes:  {\bf JA}pan - {\bf D}eutschland - {\bf E}ngland.

The proposal was signed by 48 authors, including 11 graduate students.
The abstract provides a concise summary of the relevant key features
and scientific goals of the experiment:

{\it
We propose an experiment which will be ready to take data from the first colliding beams at 
PETRA.
The apparatus consists essentially of a system of cylindrical drift chambers placed inside a thin
normal conducting solenoid.
A modular array of leadglass shower counters surrounds the coil and beyond this we have
a muon filter system.
A small angle double tagging system is provided as a luminosity monitor and for two photon physics.

Using this apparatus we intend initially to explore the following physics:

\begin{enumerate}
\item Total annihilation cross section $\epem \rightarrow$ hadrons.
\item Search for new particles by detecting weak decays to e$^\pm$ or $\mu ^\pm$
\item Check for QED processes at high momentum transfers and search for neutral weak current
effects in $\epem \rightarrow \mu^+\mu^-$ and $\epem \rightarrow hadrons$.
\item Study of hadronic final states.
\item Survey of two photon initiated reactions.
\end{enumerate}
}

Rolf Felst was elected as spokesperson of the JADE collaboration - a position he held 
until the end of the experiment in 1986.
The representatives of the groups signing the proposal were Robin Marshall (UK groups), Wulfrin Bartel 
(DESY and Hamburg University),
Joachim Heintze (Heidelberg) and Shoji Orito (Tokyo). 
In October 1976, the proposal was recommended for approval by the PETRA Research Council, PRC.

The participation of Daresbury Lab later changed over to the 
Appleton High Energy Laboratory (now Rutherford Appleton Laboratory).
In  spring 1984, the group around Gus Zorn from University of Maryland, USA,
joined the collaboration\footnote{
At that time, the \oq {\bf A}" in \oq JADE" was unofficially redirected to the new
{\bf A}merican group, such that each collaborating nation was still represented
by one letter of the collaboration's name.}.

The timeline of the JADE experiment, from its planning in early 1976 to its
shutdown at the end of 1986, and its revival in the late 1990s and beyond, is summarised in 
Table~\ref{tab:2}.

\subsection{Set-Up}
\label{sec:execution}

\begin{figure}[htb]
  \includegraphics[width=\textwidth]{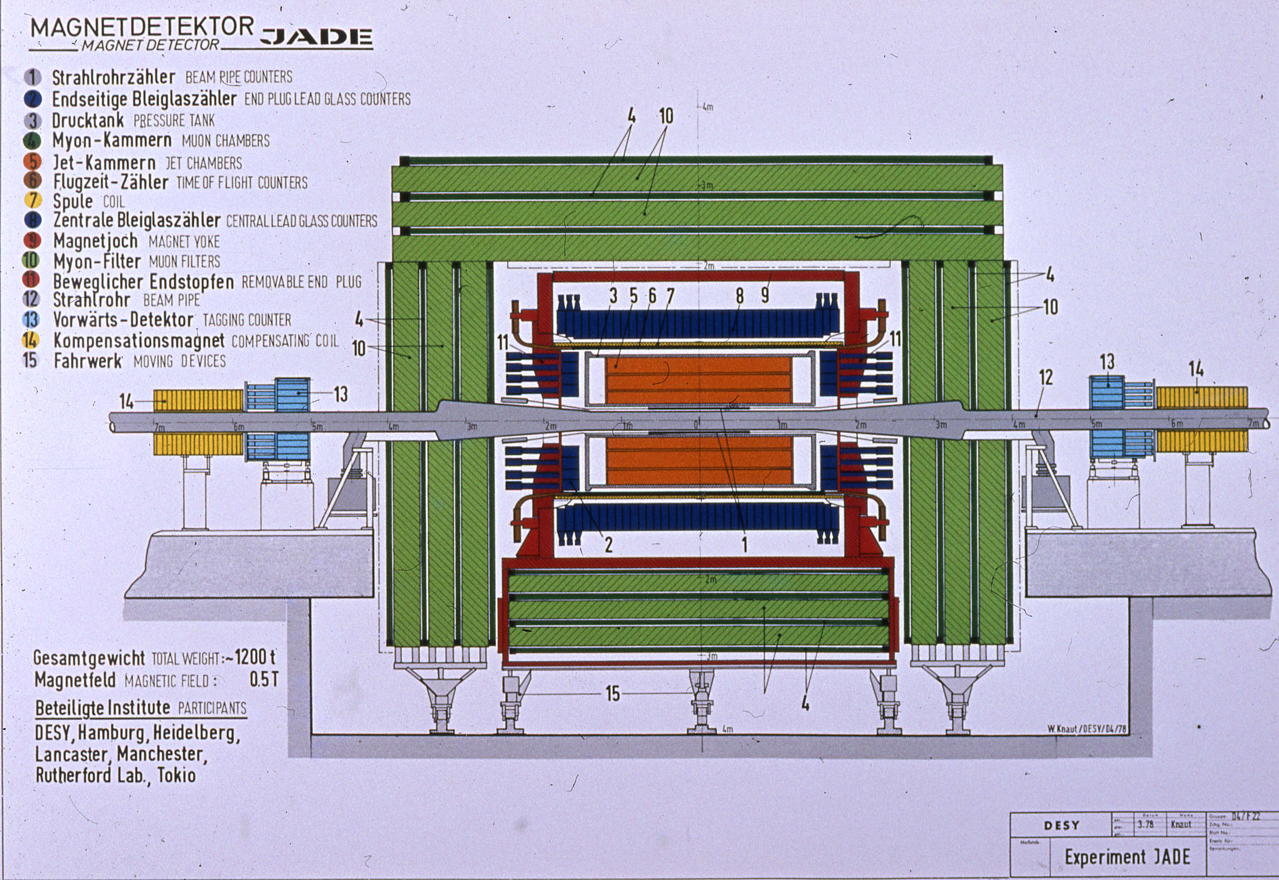}
\caption{The JADE detector in its 1979 configuration.}
\label{fig:detector}       
\end{figure}

Construction of the detector started right  after approval of the project in October 1976.
Only 2.5 years later, in January 1979, the JADE detector, fully completed, was moved into the
beam in the North-West hall of  PETRA . 
This time span seems amazingly short from today's perspective, especially as
the compact detector system - about 8 m long, 6.5 m high and 7 m wide, see Fig.~\ref{fig:detector} - exhibited features and technologies that were novel and trendsetting at that time.

\begin{table}[htb]
\centering
\caption{Timeline of the JADE experiment at the $e^+e^-$ storage ring PETRA
(PRC: PETRA Research Council; FADC: Flash Analogue-to-Digital Converter).}
\label{tab:2}      
\begin{tabular}{c | r l l}
\hline\noalign{\smallskip}
Phase & Date & & Event  \\
\noalign{\smallskip}\hline\noalign{\smallskip}
planning & 1976 & March 1-5 & discussion meeting on PETRA experiments, Frascati (Italy)\\
and &  & August & JADE proposal submitted to DESY management \\
construction & & October & JADE proposal recommended for approval by PRC \\
\noalign{\smallskip}\hline\noalign{\smallskip}
  & 1979 & February & JADE detector moved into beam \\
  & & March & first data recorded (single $e^-$-beam, $E_{beam} = 11.3$ GeV) \\
installation  & & March 26 & fatal beam loss and damage of tracking chamber \\
and  & & April 1 & removal of detector from the beam and tracker repair \\
start-up &  & June & reinstallation and data taking \\
  & & July 31 & very first results presented to PRC by J. Heintze \\
  & & August 23 & first results and evidence for the observation of the gluon \\
  & & & presented at $9^{th}$ Int. Symp. on Lepton/Photon\\
  & & &  Interactions at High Energies, Batavia, by S. Orito \\
\noalign{\smallskip}\hline\noalign{\smallskip}
  & 1981 & & mini-beta quadrupoles and reconfigured forward detectors \\
upgrades  & 1982 & & new lead-glass counters in central barrel \\
& 1984 & & installation of vertex chamber and z-chamber \\
& 1985 & & FADC upgrade of jetchamber read-out \\
\noalign{\smallskip}\hline\noalign{\smallskip}
shutdown & 1986 & Nov. 3 & shutdown of PETRA and the JADE experiment\\
& 1991 &  & phasing-out of JADE data analyses \\
\noalign{\smallskip}\hline\noalign{\smallskip}
 &1996 & & data sets rescued and copied to modern data carriers \\
data & 1997 &  & revival of JADE software  \\
preservation &1997 &- 2012  & new data analyses and scientific publications \\
and & 2009 & August 22 & JADE revival and collaboration meeting \\
revival & 2011 &  & digitisation and archiving of shift crew logbooks (1979 \\
& & & -1986), collaboration meeting protocols and internal notes \\
& 2022 &  & release of JADE data and software for public access  and\\
& & & usage, in the framework of the CERN Open Data portal. \\
\noalign{\smallskip}\hline
\end{tabular}
\end{table}

\subsubsection{The JADE Detector}
\label{sec:detector}
The detector covered 97 \% and 90 \% of the full solid angle for detecting charged particles and 
photons, respectively \cite{jade-total-x}.
A 3.5 m long solenoidal coil, 2 m in diameter and with a 7 cm thin aluminum conductor, 
produced a magnetic field of 0.5 Tesla parallel to the electron/positron beam axis.
An array of 24 scintillators surrounded the beam pipe, and 42 time-of-flight-counters were mounted
immediately inside the coil. 

A cylindrical, novel type of drift chamber, the \oq jet chamber" \cite{jetchamber}, 
was located inside the magnet volume, to accurately measure the tracks of charged particles
that emerge from the $\epem$ interaction point in the beam pipe and centre of the detector.
A cylindrical array of lead-glass shower counters was arranged outside of the magnet coil, 
consisting of 2520 lead glass blocks, individually calibrated, reaching an energy resolution
for large angle Bhabha events ($\epem \rightarrow \epem$) of $\sigma_E / E = 4 \% / \sqrt{E} + 1.5 \%$.

Two flat iron endcaps with 
96 lead glass blocks on each inner side served as part of the magnet return yoke,
closed the cylinder and completed
the lead glass system to a total coverage of 90\% of the solid angle. 
The magnet return yoke surrounded the lead glass system and
also served as the first layer of the muon filter, followed by further layers 
consisting of iron loaded concrete, interspersed by 4 or 5 layers of drift chambers 
covering 92\% of the solid angle.

Two small detectors, consisting of scintillators, drift chambers and lead-glass modules, recorded 
electrons and positrons at small polar angles (measured with respect to the direction of the incoming positron beam), providing online measurement of the luminosity of the colliding beams, and tagging 
two-photon processes ($\epem \rightarrow \epem +$ hadrons).
For a concise description of all components of the JADE detector, see also \cite{naroska}.

\begin{figure}[htb]
  \includegraphics[width=\textwidth]{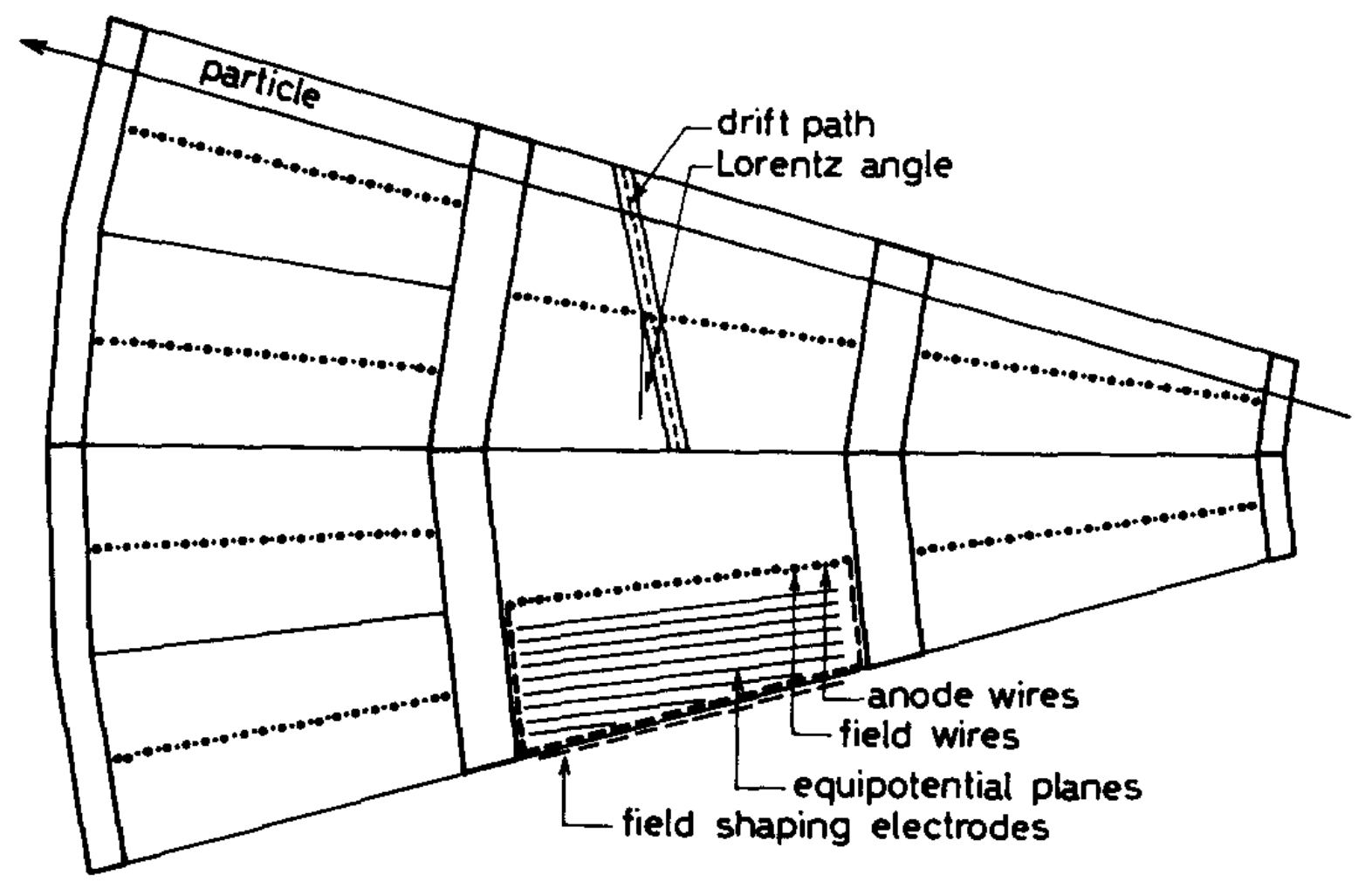}
\caption{Cross section through two of the 24 sectors of the jet-chamber.}
\label{fig:jetch-segment}       
\end{figure}

The jet chamber, together with the lead-glass calorimeter, was the key component enabling
JADE's scientific success.
Developed, constructed and commissioned by the Heidelberg group around Heintze,
it consisted of 96 drift cells, arranged in 3 concentric rings around and parallel to the beampipe,
each containing 16 sense wires and 19 potential wires plus a system of field generating electrodes
defining linear drift spaces up to 8 cm long, see Fig.~\ref{fig:jetch-segment}.
The chambers were contained in a pressure vessel, 3m long and 1.8m in diameter, filled with a gas
mixture of Argon, Methane and Isobutane at 4 atm.

Charged particles, emerging from the beam collision point and traversing the jet chamber, 
ionise the gas along their curved\footnote{through the magnetic field} trajectory, leading to a 3-dimensional measurement of up to 48 space-points for each track.
The $z$-coordinate along the beam direction was determined by charge division,
the ratio of signal amplitudes read-out on both ends of the sense wire; 
the radial distance $r$ to the beam line was given by the location of the sense wire,
corrected for effects of the Lorentz force on the drifting electrons;
the azimuthal position
$\phi$ in the plane perpendicular to the beamline was given by the drift-time and hence the distance to the 
sense wire, whereby the left-right ambiguity was resolved by alternating displacements of the wire
positions by $\pm 150 \mu $m.
The spacial resolutions obtained were 
$\sigma_{r\phi} = 180 \mu$m ($110 \mu$m after the electronics upgrade to Flash ADCs, see 
Section~\ref{sec:upgrade}) 
and $\sigma_z = 16 -32$ mm.

\begin{figure}[htb]
  \includegraphics[width=\textwidth]{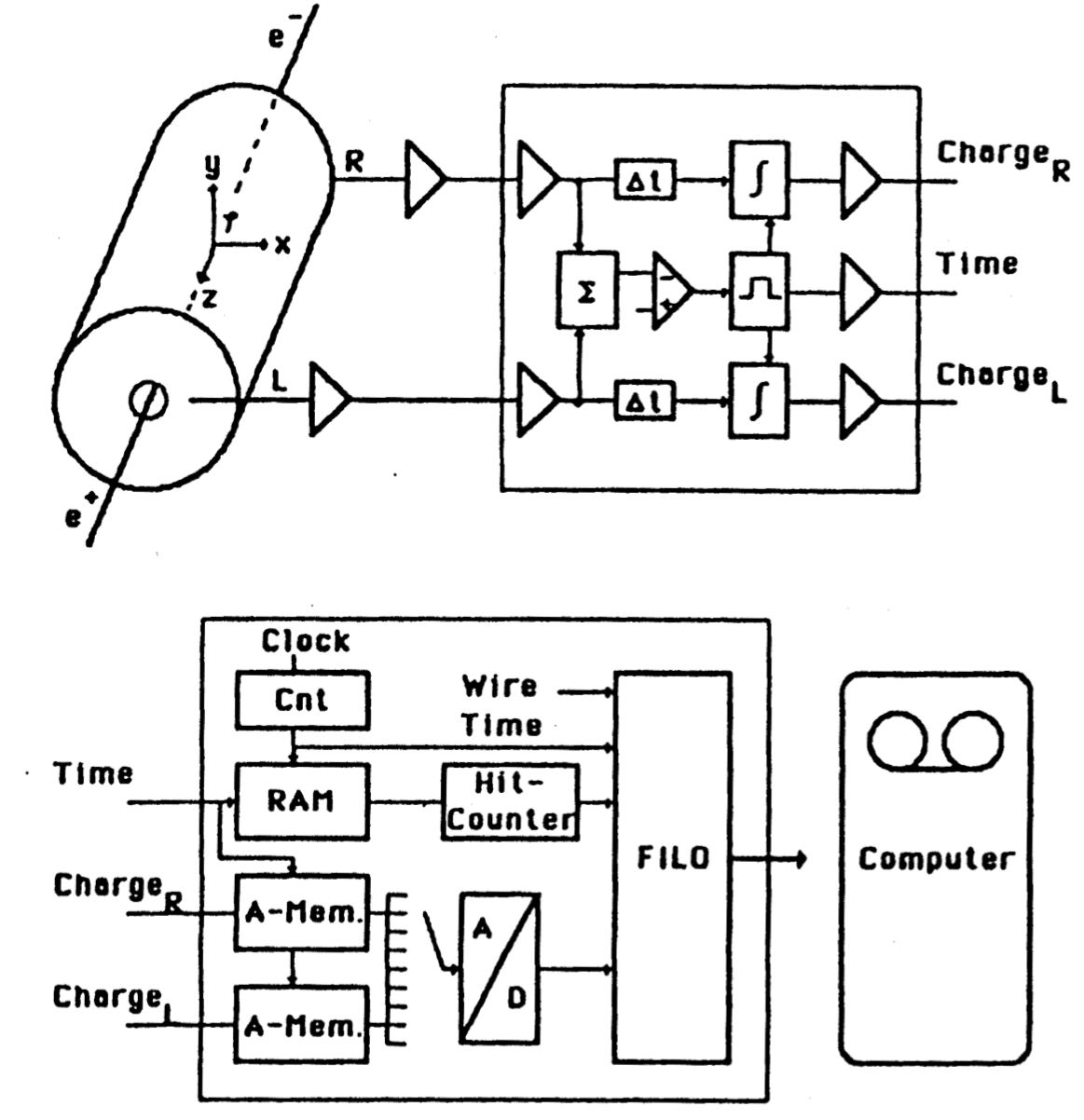}
\caption{Multihit electronics of the JADE jet chamber with analog pulse discrimination and
charge integration.}
\label{fig:jetch-readout}       
\end{figure}

The electronics for  readout and digitisation of data from the jet chamber also was custom-made,
developed and built by the Heidelberg group \cite{jetch-electronics},
comprising pre-amplifiers at both ends of each sense wire, sitting on the outer side of the
end flanges of the pressure vessel, NIM-type\footnote{Nuclear Instrumentation Module (NIM) standard, defining mechanical and electrical specifications for electronics modules used in experimental particle and nuclear physics\cite{wikipedia}.}
amplifier/discriminator/integrators generating 
pulse height \oq right", pulse height \oq left" and time signals for each sense wire, followed by 
CAMAC-type\footnote{Computer-Aided Measurement And Control (CAMAC), 
a modular crate electronics and bus standard for data acquisition and control \cite{wikipedia}.}
time and amplitude digitisers with multiple hit capacity, see Fig.~\ref{fig:jetch-readout}. 
The system was able to process up to 8 hits per signal wire with a double pulse resolution 
down to 70~ns, corresponding to  3.5~mm in space at a drift velocity of 5~cm/$\mu$s.
The charge measurement was performed using gated integrators, fast analog memories and a 12-bit 
multiplexer-ADC system (one per 8 wires). 
With this system, three-dimensional track coordinates and the energy loss of the particles passing the 
detector could be obtained.

Both the concept and realisation of the jet chamber and its associated multi-hit readout electronics 
were novel developments at that time,  generating high expectations for its
capabilities and performance. 
While this concept raised critical scepticism in the early phase of planning and approval, 
it contributed in a major way to the scientific success of JADE.

\subsubsection{Event Selection and Data Acquisition}
The two electron- and two counter-rotating positron-bunches crossed each other, at each 
of the four interaction points of PETRA, every 4 $\mu$s, i.e. at a rate of 250 kHz.
The JADE trigger system reduced this rate to a few collision events per second which were
read out and stored for physics analyses.

\begin{figure}[htb]
  \includegraphics[width=\textwidth]{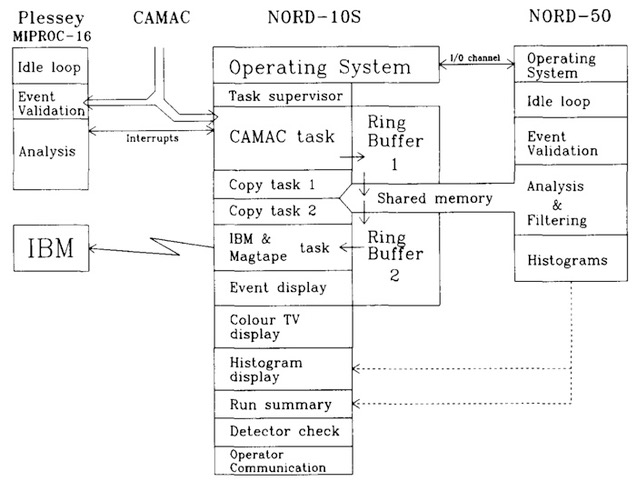}
\caption{The JADE data acqusition system (JDAS) \cite{mills}.}
\label{fig:jdas}       
\end{figure}

The trigger was organised in three levels, T1 to T3. 
The first level (T1) was based on fast analog signals from the scintillators 
and the lead glass counters, which were available about 350 ns after the beam crossing.
Based on certain requirements of configurations or energy sums, collision events were immediately
rejected, or postponed to the next level (T2), or immediately accepted for read-out.
Rejection at this stage allowed to reset the trigger and read-out system, without causing dead-time. 

The level 2 trigger was based on the data of the tracking chamber and a fast track-finding logic,
which were available about 2 $\mu$s after beam-crossing. 
Until 1982, rejection by T2 caused the loss of the next bunch crossing because of the
time needed to reset the jet chamber electronics.
From 1982, this was changed such that a T2-reject did not generate extra dead-time.

Finally, trigger level 3 used signals from the muon counters which were available about 5 $\mu$s
after beam-crossing.
Events accepted at any stage of T1 to T3 initiated an interrupt signal  to the online computers which 
read out the digitised event information.

The average amount of data read out for a typical $\epem$ collision event was 200 to 3500 16-bit words, occuring at a trigger rate between 2 and 6 Hz, depending on the PETRA beam conditions.
Multihadronic events, at a rate of a few per hour, typically comprised 4000 to 8000 words.
In the initial phase of the experiment, the data were written on magnetic tapes, which were then 
carried over to the DESY main frame computer centre for further processing and data analysis.
Later, data were directly transferred to the computing centre using a link between 
the JADE online computer and DESY's IBM main frame.

Efficient operation and control of a complex detector like JADE required a maximum of on-line computing power, memory and data handling capacity.
At the time of JADE operation, computing power as well as memory capacity was very limited, 
many orders of magnitude less than available today.
The choice made for the JADE experiment was a NORD-10s/50 dual processor from Norsk-Data, Norway, and a Plessey MIPROC-16 microprocessor.

The NORD-10s was a general purpose  
16-bit minicomputer for time-sharing applications and real-time multiprogram systems.
The CPU consisted of a total of 24 printed circuit boards, and its speed was 260 ns per micro-instruction
\cite{nord10}, corresponding to a CPU clock speed of 3.8 MHz.
The memory management system  allowed a 16-bit virtual word address to be mapped into an 18-bit physical address, so that the maximum memory available to a user program was 128 kbytes and the maximum physical memory was 512 kbytes. In addition the memory management system swapped 2 kbyte pages to and from a 66 Mbyte disc, thereby extending the memory size \oq almost indefinitely" \cite{daq}.

The NORD-10s was equipped with two 66 Mbyte disc units, two floppy-disc units, one card reader, a 1600 b.p.i. magnetic tape unit, ten terminal drivers and two external bus drivers (one for CAMAC I/O and the other for a crate that provided the link to the DESY IBM mainframe computer), a colour TV screen and a Gould black-and-white printer/plotter.

The NORD-50 was a 32-bit single-program computer which was about a factor of 3 faster than the
NORD-10. 
It controlled no peripherals at all and needed a NORD-10 which drove it via a set of registers. The particular advantage of the NORD-10s/50 system was that the two processors could access a common part of a multiport memory and were not simply linked by I/O channels. 

For digitising the data and controlling the electronics for the JADE experiment, 40 CAMAC crates were interfaced to the NORD-10s. 
The MIPROC-16 was used for part of the online event filtering scheme. 
The NORD-50 was used for a single program which performed event validation,
analysis and monitoring, including histogram filling \cite{mills}.
Several microprocessors operated within individual crates and for various purposes.
The organisation of the tasks and buffers of the JADE Data Acqusition System (JDAS) is 
shown in Fig.~\ref{fig:jdas}.

\subsubsection{Offline Computing, Data Analysis and Monte Carlo generation in a time before internet}
At the end of the 1970s, the central computer at DESY was 
based on a system of two IBM 370-168 
mainframes that were installed in 1975, each with a main core memory of 3 MByte\footnote{
The main memory was already based on silicon technology, in contrast to 
magnetic-core memory that was the predominant form of random-access computer memory before 1975.
1975 also marked the transition from punch-card programming to a terminal based 
time-sharing system (TSO). This transition happened slowly and was completed  at DESY by 1979.}.
The system was upgraded in 1979 by the purchase of an additional mainframe IBM 370-3033,
which basically doubled the total available CPU power.
The 3033 provided a cycle time of 58 ns, equivalent to a CPU clock speed of 17.2 MHz.

During data taking, the JADE raw data were transferred to 
the DESY computing centre using a fast internal link, and
written on machine-room tapes with a capacity of 160 MB each (\oq REFORM" data sets). 
These data were then further processed, calibrated, reduced, and pre-analysed off-line
on the IBM mainframe, and the resulting data files, called \oq REDUCn", with n=1 being the
main stream for all selected data, and n=2 for a special selection with slightly varied selection
criteria better adopted to the needs of two-photon physics, were again written to machine room tapes.

Due to the absence of the internet or any of its precursors, the data basically resided at DESY, 
and physics analyses proceeded mainly on the DESY IBM mainframe.
Also the generation of Monte-Carlo based event samples, required for the study and correction for
limited detector response and resolution, was performed on the DESY IBM mainframe.
As all the PETRA experiments processed their data in a similar way,
the DESY computing system  was almost always operated and used at its limits,
and there was need for significantly more computing resources than available at DESY,
e.g. for reprocessing of all data sets, or for sufficiently large statistics and variants of MC production.
This was
a major bottleneck requiring innovative, farsighted and collegial usage of the available computing 
resources.

There were, however, occasions of accessing and using significant additional off-line computing
resources for JADE.
The Tokyo group of JADE faced the problem that many of their students needed to perform and finalise
their data analyses back at Tokyo, due to limited resources for student's travels and staying abroad.
The group therefore acquired and installed a FACOM M190 mainframe system, a clone of and thus compatible to the IBM 370, at their lab at Tokyo.
That computer was equipped with a copy of the DESY software system, such that
the JADE software and all computing jobs at this machine ran exactly like at DESY,
without further modifications and adjustments.
Being well-equipped with local computing power, 
this also allowed for additional reprocessing runs of the data, and for extensive MC event 
generation. 
The in- and output-data, however, always had to be transferred by sending large batches of magnetic
tapes (c.f. Fig.~\ref{fig:tapes}), between Tokyo and DESY.

Another occasion for using significant extra off-line computing resources was procured 
by the Heidelberg group:
The mainframe computer of the University of Heidelberg, an IBM~3081 - a 
successor of the IBM 370 series - at that time was routinely switched off during week-ends,
due to low user demands and also for saving resources. 
In 1983, the JADE group at Heidelberg succeeded to obtain permission for exclusive usage of the
entire computing centre over week-ends, free of charge, but supplying its
own team of trained (student) operators for that time, for mounting in- and output tapes but
also for assuring smooth operation of the hard- and software systems, and for switching on and off the computer centre for these special weekend operations\footnote{
The installation of the JADE (and DESY) software, however, was difficult, due to a different
system software on the 3081, and finally was
not successfully completed.
Instead, load-modules of programs being produced on the DESY mainframe proofed
to run without problems on the 3081, such that productions and generations that did not
need much user intervention and adjustments could formidably run on the 3081.
Monitoring and steering of this production at Heidelberg was performed using
an early version IBM-PC at DESY, with an acoustic-coupled phone connection
to the Heidelberg mainframe with 800 baud transfer speed.
However, the transfer of in- and output data, and of the load modules, had to be arranged
via large magnetic tapes, being transported by scientists travelling between Hamburg
and Heidelberg by train.}.

\subsection{Operation and Detector Upgrades}
\label{sec:operation}

In spite of the very short time for planning, construction and commissioning, the 
detector  performed as expected right after
its completion and installation in February 1979.

\subsubsection{Early Set-Backs}
\label{sec:disasters}

However, already on March 26 1979, a fatal loss of the PETRA beam right into the detector 
resulted in major damage of the jet chamber, 
c.f. Table~\ref{tab:2}.
The beam loss was caused by a trip of part the magnet system due to a failure of the cooling water supply.
The electrons of the PETRA beam hit the beampipe close to the JADE detector, producing 
showers of energetic particles traversing the jet chamber, ionising the
gas and causing major distortions of the electric field in the drift chamber. 
This in turn lead to large oscillations of the anode signal wires, some of which would contact the 
neighbouring cathode wires.
Due to a too large capacity of the decoupling capacitors in the high-voltage supply for the electric fields 
inside the chamber, some of the 50~$\mu$m thin sense wires 
therefore melted when making contact with the cathode wires.

The scientists on duty at this night shift of March 26 analysed the situation 
and found that in more than half of the jet chamber cells at least one wire 
was broken and that these cells could no longer be operated.\footnote{
They must have been shocked as the only thing they noted in the experiment's log book was: 
\textit{6:40 Strahl verloren (beam lost) END.}}

Rapid decisions and
actions were taken in order to repair the damaged jet chamber.
On April $1^{st}$ 1979, within 24 hours the JADE detector was rolled 
out of the beam and the beam vacuum was reestablished. 
The pressure vessel with the jet chamber was removed from the surrounding detector, 
carefully packed into a large truck and driven back to the Physics Institute of University
of Heidelberg, where the jet chamber had been built and completed just few months before.
Disassembly of the chamber, its repair, re-assembly, transport back
to DESY, re-installion and commissioning at DESY took little more than 2 months,
such that the JADE detector started to take collision data again
in June 1979.

JADE had to catch up with its competitor experiments at PETRA, MARK-J, TASSO and PLUTO,
who took their first data during JADE's repair shutdown.
The very first results obtained in this previously 
unexplored energy range showed that the total hadronic cross section, 
a measure for the number of quark generations, showed no evidence for the hypothetical top-quark.
However, these early data already exhibited first signs 
for $the$ major discovery to be made at PETRA, the gluon - see Section \ref{sec:3} below.

Data taking of JADE started right after re-installation and commissioning of the
detector, and 
very first analysis results of these early JADE data were already presented at the 
PRC\footnote{PETRA Research Committee; 
later generalised to Physics Research Committee}
meeting on July 31 1979 by Joachim Heintze,
and at the $9^{th}$ International Symposium on Lepton-Photon Interactions at High Energies, held at
Fermilab / Batavia (USA) in August 23-28, 1979, in a talk given by Shoji Orito \cite{orito}. 

In the course of 1980 running, a total of 9 out of the 96 drift cells of the central jet chamber
developed HV-shortages, so that these had to be taken out of HV-supply and
data-readout (\oq dead cells").
As a consequence, the detector was again opened and the jet chamber vessel removed and
driven to Heidelberg for repair, in the winter shutdown of PETRA of 1980/1981.
The repair work was completed during the scheduled shutdown.\footnote{
The action was risky due to deep-winter conditions 
on the way to and from Heidelberg,
two 600 km drives through deep snow and the need to keep the detector at temperatures safely above
$15{\textdegree}$C at all times.
In addition, it was a challenge to manage disassembly, repair, assembly and
commissioning during the short available time over the holidays.
Being experienced with this kind of emergency actions, the Heidelberg crew succeeded to
deliver and re-install a fully functional jet chamber in time 
again.}
%
\begin{figure}[htb]
  \includegraphics[width=\textwidth]{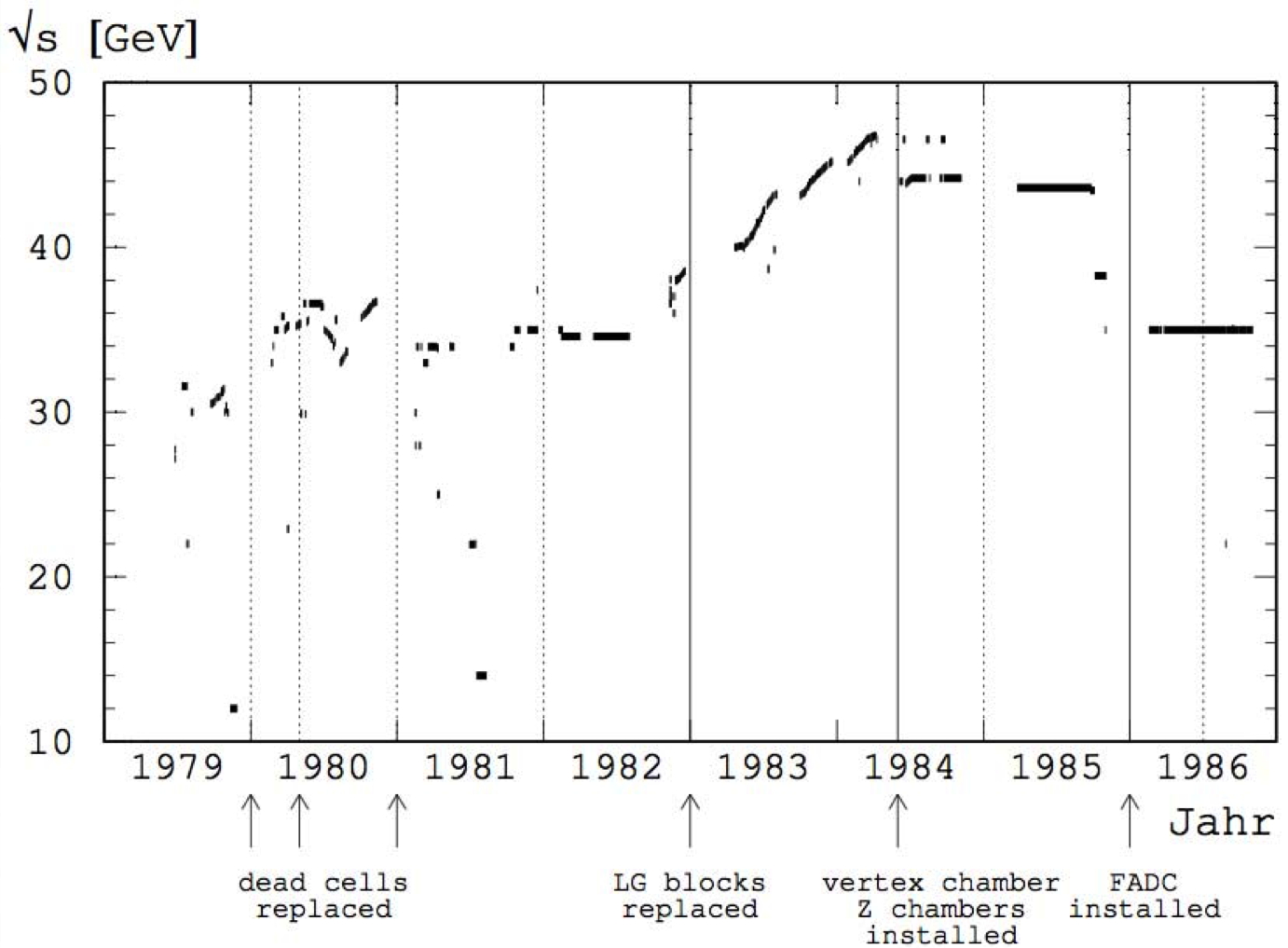}
\caption{$\epem$ centre of mass energies of data taken as function of time.
Major repairs and upgrades of the JADE detector are indicated.}
\label{fig:energysum}       
\end{figure}
\subsubsection{Routine Running}
\label{sec:running}
\begin{figure}[htb]
  \includegraphics[width=\textwidth]{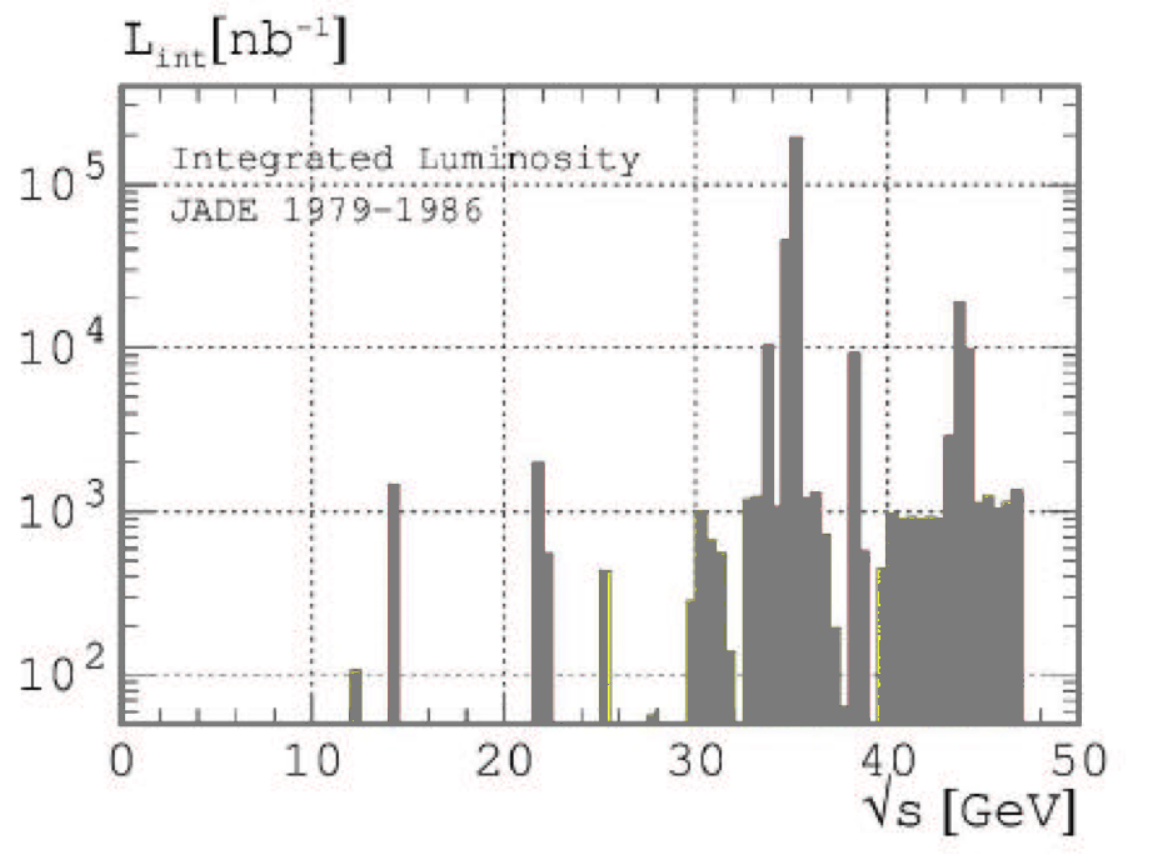}
\caption{Integrated luminosity collected by JADE as a function of the centre of mass energy.}
\label{fig:lumisum}       
\end{figure}

The JADE detector recorded $\epem$ collision data from June 1979 to November 1986, when
PETRA was finally shut down.
The variation of the centre-of-mass energy  $\sqrt{s}$ as a function of time is
summarised in Fig.~\ref{fig:energysum}, and
the integrated luminosity recorded at the different energies is given in Fig.~\ref{fig:lumisum}.
The total integrated luminosity summed up to 216 pb$^{-1}$.

In the first three years of running, PETRA and the experiments operated in 24/7 mode for
typically few weeks in a row, intercepted by short periods of shut-down for maintenance
and upgrades of the accelerator and the detectors.
During the last years of operation, extended data runs lasting several months 
became common routine.

Detector operation during data-taking 
required two persons being on shift in the JADE control room, controlling and steering
the performance of detector hardware, data taking and data flow, performing 
small repairs and fixes, 
informing and calling experts in case of major problems and happenings, and
coordinating with the other experiments and the PETRA control room (PKR),

Shifts were arranged in three 8-hours shifts throughout the day, from 0:00 to 8:00,
from 8:00 to 16:00 and from 16:00 o'clock to midnight.
In addition, one expert for each of the major detector subsystems was
assigned to be on-call for cases of major problems, typically for periods of 24 hours,
adding up to a total of 12-15 people being in charge and responsible for operation
of JADE every day.
During times of data taking, these were major challenges
for a collaboration consisting, at any given time between 1979 and 1986,
of 50 to 60 people, who partly were based at their home institutions rather
than at DESY. 

\begin{figure}[htb]
  \includegraphics[width=\textwidth]{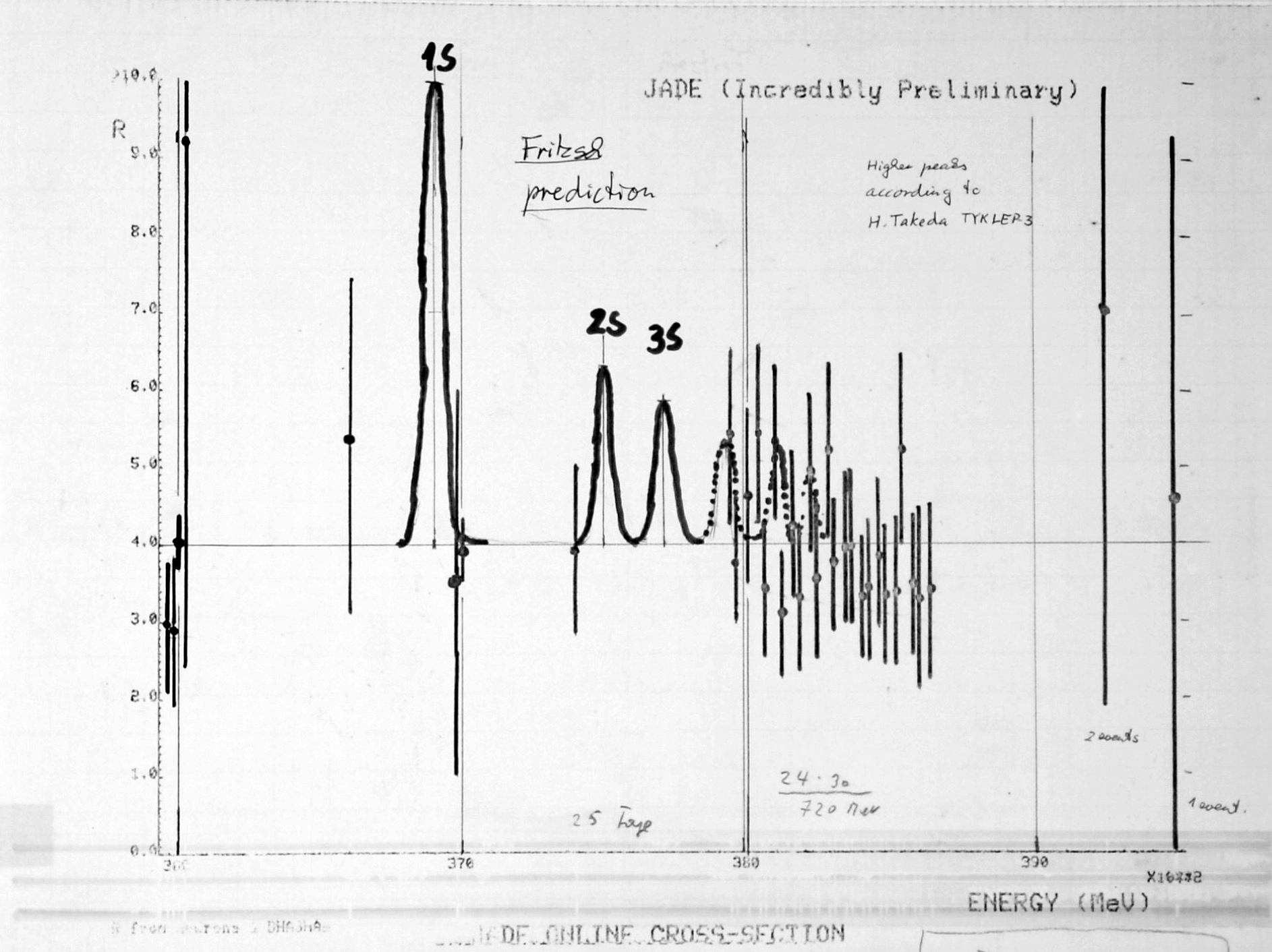}
\caption{Online version of the \oq R-plot", the normalised hadronic cross section
as function of the c.m. energy, from 36 to 39.5~GeV,
as plotted on 21 December, at the end of the 1982  
energy scan to search for the top quark
(JADE online logbook \#10, page 73 \cite{minutes}). 
Drawn-in by hand is a sketch of the expectation for the ground state and higher 
$t\overline{t}$ resonances
according to a then actual theoretical prediction.
The expectation for $R$ in case of 5 quark flavors, without the top-quark, is 3.8,
and for 6 quarks, including the top-quark in the continuum, about 5.2.
The horizontal line at $R = 4$ is drawn to guide the eye.}
\label{fig:r-online}       
\end{figure}

The first data collected at run start in 1979 were at c.m. energies around 30~GeV and concentrated 
around 35~GeV  in the course of the end of 1979 to late 1982, with short interceptions at 22 and at 
14~GeV in 1981.

From late 1982 to the middle of 1984, the c.m. energy of PETRA was gradually increased,  
in steps of 30~MeV, from 35~GeV up to PETRA's maximum achieved energy of 46.78 GeV.
At each energy point, about 200nb$^{-1}$ were accumulated, on average, for each of the four
experiments.
This was the exciting time of scanning for the top-quark, the predicted but yet undiscovered $6^{th}$
quark, whose mass was expected, at that time, to be about 20~GeV, and the $t\overline{t}$-ground state to have a mass between 36.8 and 38 GeV, see e.g. \cite{fritzsch83}.\footnote{
In order to demonstrate the high levels of expectation and of motivation at that time, we 
reproduce some of the comments made and documented in minutes of the weekly JADE meetings, and in the online logbooks (scanned versions available at \cite{minutes}):

\textit{JADE wants to find top (be-)for Xmas!} (minutes of JADE meeting on 18 November 1982)

\textit{Main worry is now that we are running already above $t\overline{t}$ ground state, since 
Harald Fritzsch predicts $M^{1s}_{t\overline{t}} = 36,9 \pm 0.1~GeV$. 
The top scan reached 38.4~GeV - Oh boy!} (minutes of JADE meeting on 16 December1982)}

A just-for-fun plot of the normalised total hadronic cross section, produced on-line
on 21 December 1982, at the end of the 1982 energy scan, stopping at 39,5~GeV 
c.m. energy, including hand-drawn expectations for the lowest $t\overline{t}$ bound states 
according to \cite{fritzsch83}, is reproduced in Fig.~\ref{fig:r-online}.\footnote{
\textit{Hurrah - das top ist daaaa! Und morgen kommt der Minister} (engl.: Hurrah -
the top (quark) is here! And tomorrow the minister will come)
was another
outbreak of excitement, in April 1984, noted in the online logbook, based on
a (statistically not really significant) fluctuation of $R$ observed at the highest PETRA
energy of 46,78~GeV.}

After the disillusion of not finding the top-quark by the middle of 1984, the 
PETRA experiments and the PRC
decided to run at c.m. energies around 44~GeV 
for the rest of 1984 and the entire 1985, 
avoiding the large drop of machine luminosity at the highest
achievable energies reached during the scan.
In order to significantly test QCD and its running coupling strength,
JADE proposed to also collect high statistics at lower c.m. energies like 22~GeV, 
where only a limited amount of data had been collected in 1981.
Instead, the decision was to operate PETRA at 35~GeV during its final year of 1986,
to make use of the best performance and highest luminosity, thus collecting 
a maximum of data.

\begin{figure}[htb]
  \includegraphics[width=\textwidth]{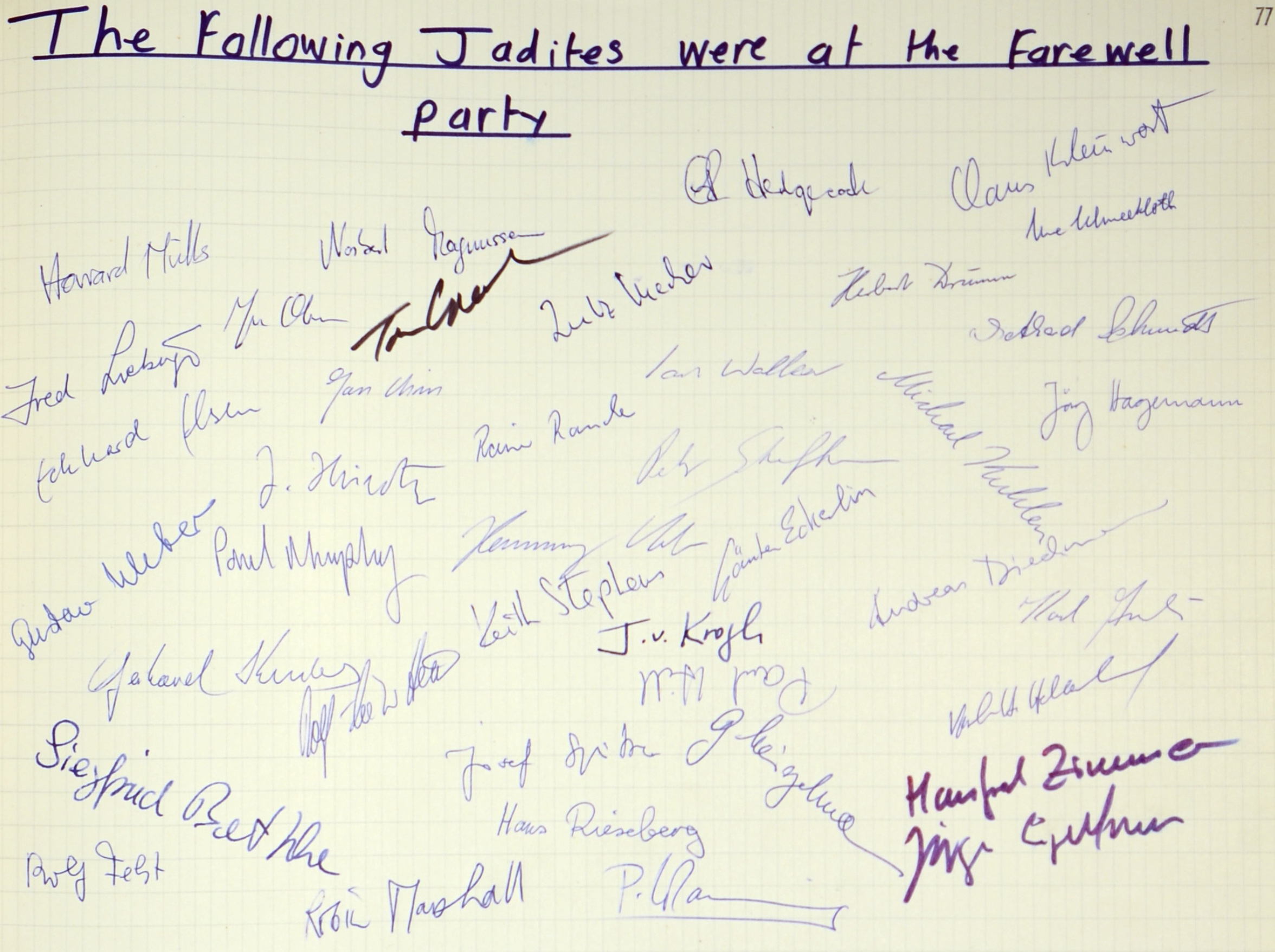}
\caption{JADE collaborators present at the farewell party on the occasion of the
shutdown of PETRA and its experiments, in the early morning of November 3, 1986
(JADE online logbook \# 21, page 77 \cite{minutes}).}
\label{fig:farewell}       
\end{figure}

PETRA and the experiments were shut down in the early morning hours of November 3, 1986,
after 8 years of successful operation.
This gave rise to the earliest JADE collaboration meeting ever, starting at 6 o'clock
in the morning, at the detector in the PETRA North-West experimental hall, with more than 40
Jadites present, see Fig.~\ref{fig:farewell}, and a formidable breakfast including a 
French Champagne endowed and served by Joachim Heintze.

\subsubsection{Detector Upgrades}
\label{sec:upgrade}
During its eight years of active lifetime, 
the JADE detector was subject to several major repairs, 
and also to a number of significant upgrades of the detector hardware.
Major disassemblies of the detector and repairs of the central jet chamber 
during the first two years of running were already 
reported in Section~\ref{sec:disasters}.
From 1981 on, the detector - in its configuration as described in
Section~\ref{sec:detector} - 
received several major upgrades, 
to improve the detector performance and modernise some technologies.

\begin{itemize}

\item
In March 1981, the beam focussing of PETRA was reconfigured to the new, so-called 
mini-beta scheme, which required the installation of new focussing quadrupoles
on both sides of and as close as possible to the beam interaction region, for all experiments.
In the case of JADE, these quadrupoles replaced the previous compensation 
magnets and the forward-detectors, c.f. items 13 and 14 
in Figure~\ref{fig:detector}.
With the new scheme, no compensation magnets 
to counter balance the effects of the JADE solenoidal magnetic field
on the beam dynamics
were necessary any more;
however new forward detectors and a new layout of the beam-pipe 
had to be designed and installed inside the JADE detector.
The stronger focussing of the mini-beta-scheme resulted in an increase of the
luminosity, i.e. the number of collision events per second, by up to a factor of three,
and thus was instrumental for the overall scientific success of the PETRA experiments.

\item
In the winter shutdown of 1982/83, the central 20\% of lead-glass counters 
in the barrel region of the electromagnetic calorimeter were replaced 
by counters of higher density (Schott SF6), exhibiting 15.7 radiation lengths 
instead of the 12.5 radiation lengths of the other (Schott SF5) barrel
counters. 

\item
In 1984, a high resolution vertex chamber with Flash-ADC (FADC) readout was installed, 
replacing the beam pipe scintillation counters
and extending the measurement of charged particle trajectories by the 
central jet chamber to smaller radii \cite{kleinworth1989}.
This upgrade of the tracking system improved the accuracy of the track extrapolation
to the main interaction point by more than a factor of 2, 
and made precise measurements of the lifetimes of b-quarks 
and $\tau$-leptons possible.

\item
Also in 1984, a new z-chamber was installed.
It was mounted onto the outer shell of the jet chamber pressure vessel,
with sense wires \oq wrapped" around the central tracking detector, and their sensitive drift direction 
parallel to the $\epem$ beam, 
providing a precise determination of the z-coordinate of tracks at large distance from the beam.
The installation of the z-chamber required a major operation to remove the jet chamber pressure
vessel and to cut it from its high-pressure gas supply.

\item
Finally, during the winter shutdown of 1985/86, only one year before the end of PETRA
$\epem$ collider operation, the initial multi-hit readout of the jet chamber
was replaced by a 100~MHz Flash-ADC system with microprocessor pulse analysis and readout
\cite{fadc,fadc2}.
This new system improved the spatial resolution of the central jet chamber 
from 170 to 110~$\mu$m, and the double track resolution from 7 to 2 mm.
This system provided the test-bed for the tracking chamber readout at the next generation of $\epem$
experiments, in particular the OPAL detector at  the LEP collider at CERN which in many respects can
be considered a successor of JADE.

\end{itemize}

%% file: Sec-03.tex
\section{Physics Highlights}
\label{sec:3}

The analysis of JADE data and publication of scientific results proceeded in two phases.
The first lasted 
from 1979 to 1991 and so came to an end about 5 years after the end of data taking.
The second phase started in 1997 with the revival of the JADE data, 
followed by more analyses, as described in Section~\ref{sec:4}.
Studies of the first phase covered a broad range of particle physics topics, predominantly 
in the fields of Strong Interactions 
and hadron production, of the Electro-Weak Standard Model, of two photon interactions, of
searches for the top-quark as well as searches for new physics beyond the Standard Model.
In total, 74 publications were issued in peer-reviewed scientific journals.

To date - more than 35 years after the end of data taking - these publications received 
more than 6800 citations, i.e. 92 citations per publication on average, and an h-index
of 44 \cite{inspirehep1}.
Two of these \cite{Bartel1986,Bethke1988}, with 970 and 653 citations, respectively, 
occupy positions number one and four of the most cited papers of all PETRA experiments.
Both these JADE papers pioneered jet physics and the experimental verification of Asymptotic Freedom
of quarks and gluons, a key-feature of Quantum Chromodynamics.

The first publications on the discovery of the gluon, 
by the TASSO  \cite{Tasso-gluon}
and the Mark-J \cite{Mark-J-gluon} collaborations, 
score second and third on the list of most cited PETRA publications, 
followed by the ones from PLUTO \cite{Pluto-gluon} 
and from JADE \cite{Jade-gluon} on positions 6 and 7, respectively.

In the following subsections, some of the physics highlights of JADE will be reviewed in more detail. 
Only those publications will be described which received more than 100 citations;
these are 14 publications out of the total of 74 by the time of writing this article.
The publications not explicitly mentioned here, however, also presented significant
and often pioneering results that advanced and shaped the field - see \cite{naroska} 
for a more complete review of the physics results of JADE.

\subsection{QCD and Jets}
\label{sec:qcd}
The theory of the Strong Interaction
between quarks and gluons,
Quantum Chromodynamics (QCD) \cite{gluons,gross,politzer}, 
was formulated just a few years before the 
proposals for the PETRA $\epem$ collider and its experiments
in 1974 to 1976, c.f. Section~\ref{sec:2}.
The most outstanding results of PETRA, the proof of the existence of the gluon as carrier 
and exchange quantum of the Strong Force, and the verification of the
force's distinct feature of Asymptotic Freedom, 
were not yet identified as central themes in any of the
proposals of PETRA and its experiments.
The general importance of hadronic final states of high energetic $\epem$ annihilations,
however, was recognised as one of the central physics themes at PETRA, and has largely influenced
the design and layout of the detectors - for example JADE's central tracking chamber, the jet-chamber,
and its sophisticated multi-track resolution capabilities.

\subsubsection{The Gluon}
\label{sec:gluon}
Topological distributions of hadrons produced in the reaction $\epem \rightarrow \ hadrons$
revealed a predominant 2-jet like structure already at
PETRA's precursor $\epem$ colliders SPEAR \cite{2-jet} and DORIS 
\cite{2-jet-pluto}, providing evidence for the underlying process of
$\epem \rightarrow q\overline{q}$ and susbsequent fragmentation of the quarks to hadrons, 
with limited transverse momenta of the hadrons w.r.t. the initial quark directions.

\begin{figure}[htb]
  \includegraphics[width=\textwidth]{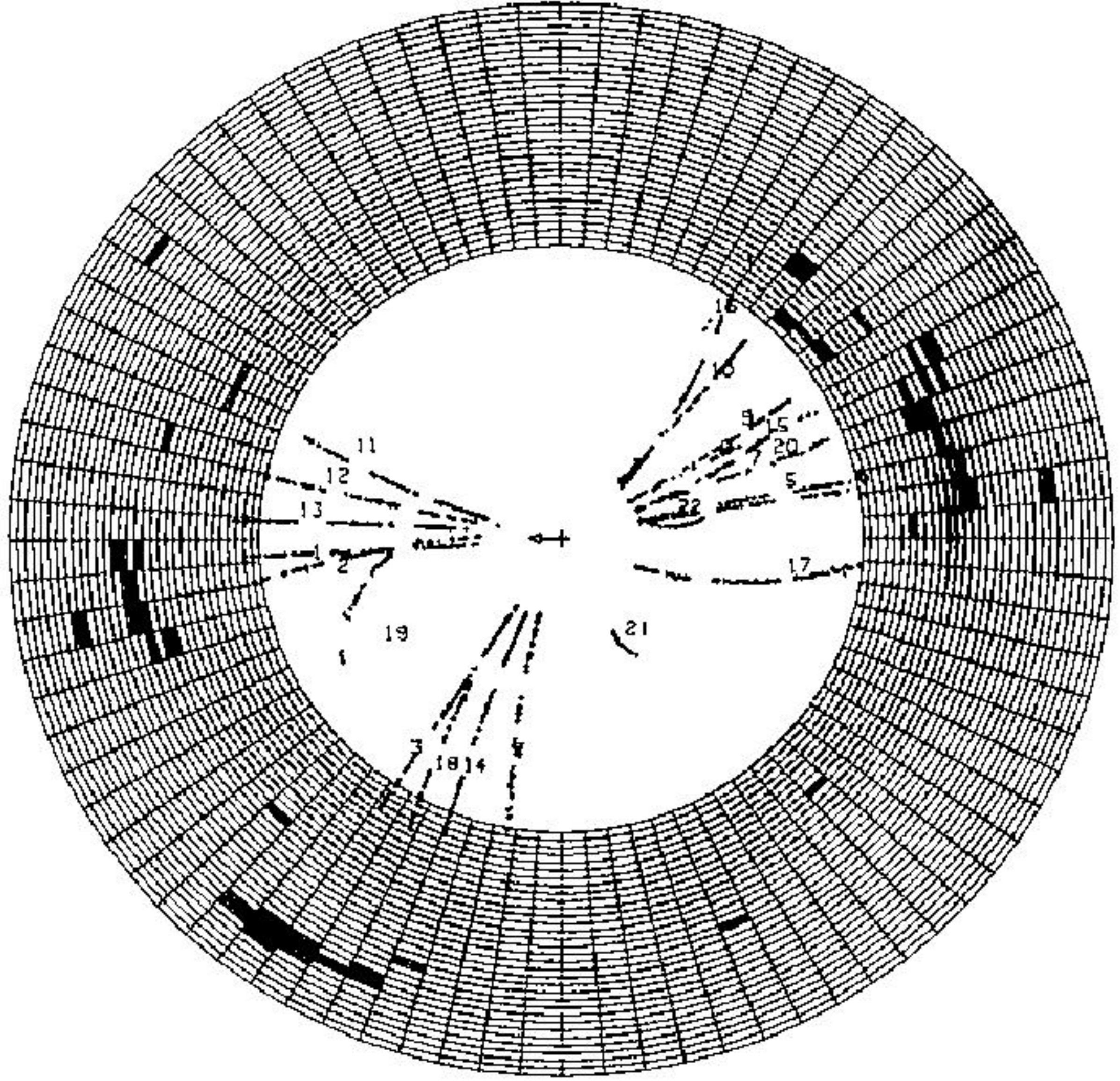}
\caption{3-jet event recorded at $E_{cm} = 33$~GeV,
displayed as projection of hits in the central Jet-chamber to the plane
perpendicular to the beam axis (central cross), and 
in a perspective view of the lead-glass counters. 
Those counters hit by particles are filled in black.}
\label{fig:3jet}       
\end{figure}

In 1976, the possibility of observing a third jet at higher c.m. energies, originating from a 
hard gluon radiated off one of the quarks,
$\epem \rightarrow q\overline{q}g$, was predicted \cite{exp-gluons}.
Gluon radiation in the context of QCD should
lead - in increasing order of the gluon's hardness\footnote{\oq Hard" in this context means
high gluon energy and large radiation angle w.r.t. the emitting quark.} - to the widening of one of the quark-jets, to a planar event configuration and, finally, to the emergence of a third jet.

Indeed, such event configurations became visible right from the start of data taking at PETRA, at 
c.m. energies around 30~GeV.
Corresponding results were presented,
first by the TASSO and then by the MARK-J, PLUTO and JADE experiments, at the 
European Physical Society (EPS) conference in Geneva and the Lepton-Photon Symposium at
Batavia in summer 1979 \cite{conf79}.
The gluon discovery publications of TASSO, MARK-J, PLUTO and JADE were submitted to 
journals in summer 1979 \cite{Tasso-gluon,Mark-J-gluon,Pluto-gluon,Jade-gluon}.

In \cite{Jade-gluon}, JADE reports about an excess of planar events in their sample of 287 hadronic
events, at a rate which cannot be explained by statistical fluctuations in the
standard two-jet process. 
The planar events, mostly consisting of a slim jet on one side and a broader jet on the other, are
shown to possess three-jet structure by demonstrating that the broader jet itself consists of two collinear jets in its own rest system.
Topological event shape distributions are compared with the predictions of model calculations
based on $q\overline{q}$ production and hadronisation with different transverse momentum 
parameters $\sigma_q$, and with a model including gluon radiation to $q\overline{q}g$ final states,
according to the expectations of QCD in leading order perturbation theory.
These studies strongly suggest gluon bremsstrahlung as the origin of the planar three-jet events. 
A first estimate of the value of the strong coupling constant resulted in 
$\alpha_s(q = 30\ GeV) = 0.17 \pm 0.04$.

The rate of radiative gluon emission is proportional to the strength of the strong coupling constant which itself is energy dependent. 
At PETRA energies the fraction of events with gluon emission is  around 10\%.
The initial evidence for 3-jet production was therefore mainly based on statistical distributions
of topological event shape distributions, and comparisons with predictions of various 
model calculations.
Nevertheless, a small number of individual hadronic events 
with visible 3-jet structure were already shown at the conferences in summer 1979.
With more statistics gathered, the significance for gluons as the origin of planar events 
with clear 3-jet structures grew.
A particluar nice, \oq golden" 3-jet event recorded by JADE in 1981 is displayed 
in Fig.~\ref{fig:3jet}.
Three well separated bundles (jets) of  tracks of charged particles are clearly visible by eye, 
directly from the coordinates of measured hits in the jet chamber, without the need for
displaying fitted track lines. 
The same structure is also visible from the clusters of hits in the lead-glass electromagnetic calorimeter.

\subsubsection{Jet Physics}
\label{sec:jets}
The dominant 2-jet structure and the observation of planar 3-jet events 
in hadronic final states at PETRA energies boosted the acceptance of 
the quark-parton model and of QCD as the theory of the Strong Interaction. 
JADE - with its high resolution central jet-chamber and lead-glass calorimeter - was
well suited to develop and advance the physics of and with hadron jets, and
to further use jets for significant tests of QCD.

\begin{figure}[htb]
\begin{center}
  \includegraphics[width=0.7\textwidth]{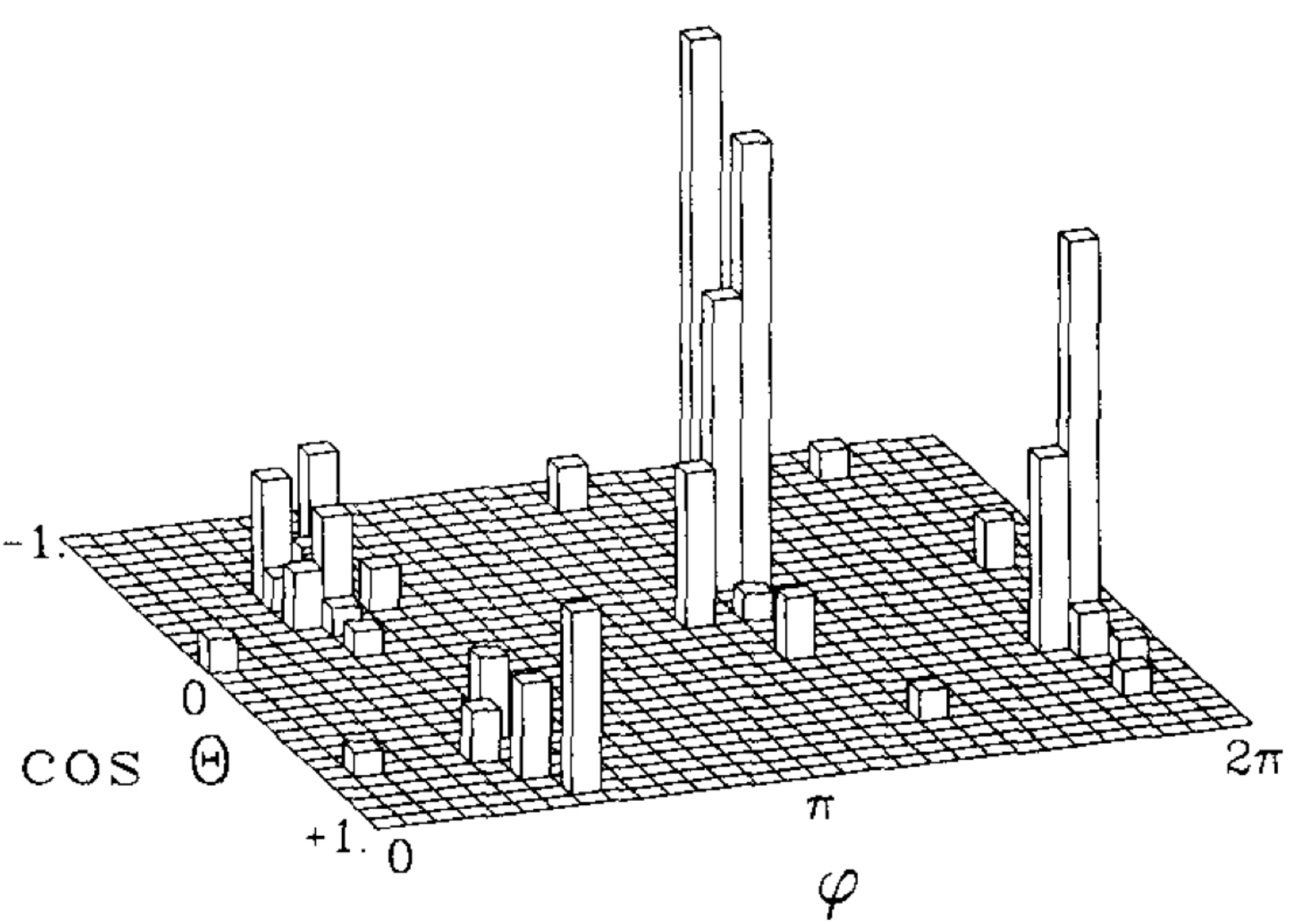}
\end{center}
\caption{Energy flow in the $\phi$ - cos$\theta$ - plane of an event measured with the JADE detector at 46.7 GeV c.m. energy.
$\phi$ is the angle in the plane perpendicular to the beamline and $\theta$ is the angle with respect to the beamline. The highest bin corresponds to an energy of 6~GeV\cite{jet-rates}.}
\label{fig:4-jet}       
\end{figure}

After establishing the existence of  gluons and of 
hard gluon radiation through the observation of 3-jet events, it was consequential 
to verify if also 4-jet events, a process  predicted by second and higher order
perturbation theory, can also be seen at PETRA.
In a study of event shape distributions that are sensitive to non-planar 
event structures 
\cite{nachtmann}, 
JADE indeed found first evidence for the
\oq Observation of Four - Jet Structure in $\epem$ Annihilation at 
$\sqrt{s}$ = 33 GeV" \cite{4-jet}.
The energy flow of an event with four well separated jets, 
later observed at the highest PETRA energies,
is displayed in Figure~\ref{fig:4-jet}.

The observed production rates of 4-jet like events, however, appeared to
be larger than predicted by the calculations and models based on 
QCD in second order ($\oaa$) perturbation theory.
This motivated to perform further and more detailed studies of jet production rates
and their dynamics, and to test QCD to a more detailed level.

For this purpose, JADE developed a jet finding algorithm \cite{jet-rates} that 
defines and reconstructs jets in terms of a resolution parameter, the minimal
scaled invariant pair-mass $y_{cut} =$\ min$(M_{ij}^2 / E_{vis}^2)$ that is allowed for
any pair of objects $i$ and  $j$ which are considered as resolvable jets,
where $E_{vis}$ is the total (visible) energy of all objects of an event\footnote{The 
algorithm starts with calculating the scaled invariant pair masses $y_{kl}$ for all pairs 
of particles, and replaces the pair with the smallest value of $y_{kl}$ by a new object
with four-momentum $p_k + p_l$. 
The procedure is repeated until all pair masses exceed the value of 
the resolution parameter $y_{cut}$.}. 
The algorithm is
infrared and collinear safe, i.e. it is insensitive to
a low-energy or a small-angle splitting of an object.
This feature avoids infrared and collinear singularities in perturbative QCD calculations,
such that theory, together with the hypothesis of \textit{local parton-hadron duality} \cite{lphd},
provides reliable and well defined predictions of the expected dynamics of hadron jets,
and thus the means for quantitative
experimental verifications of QCD.

In \oq  Experimental Studies on Multijet Production in $\epem$ Annihilation at PETRA Energies"
\cite{jet-rates}, the most highly cited publication of JADE and of all PETRA experiments,
JADE introduced this jet algorithm and published 2-, 3-, 4- and 5-jet event rates as a
function of the jet resolution parameter $y_{cut}$, in the c.m. energy range of 14.0 to 46.7~GeV.
The data were compared with model calculations \cite{lund} 
that were based on the $\oaa$ QCD
matrix element calculations of Ref. \cite{gks}\footnote{
The inclusive cross section for $\epem$
annihilation into 3- and 4-partons, including all virtual and real contributions 
in complete $\oaa$ QCD and for vanishing jet resolutions, were calculated first by Ellis, Ross and Terrano \cite{ert}. 
These matrix elements, however, could not directly be included into Monte Carlo models 
for jet fragmentation.
The issue of different variants and developments of $\oaa$ QCD calculations like those described in Refs.
\cite{gks}, \cite{GS} and \cite{KL} are further discussed e.g. in \cite{KM}.},
and with a parton shower model \cite{qcd-show-1} including multiple gluon radiation according to leading logarithmic approximations (LLA) and soft gluon coherence effects \cite{qcd-show-2}.

The measured 2- and 3-jet rates and their dependence on the size of the jet resolution were
well described by the $\oaa$ QCD model, but its deficiency in describing 4- (and higher) jet events
was further substantiated. 
The QCD shower model described 4- and 5-jet rates well but fell short of reproducing the relative rates of hard 3-jet events.
Both these observations led to further developments and optimisations of the underlying QCD
calculations and their implementation in QCD models, see e.g. \cite{bethke-scale,opal-jets}
and references quoted therein.

\begin{figure}[htb]
\begin{center}
  \includegraphics[width=0.7\textwidth]{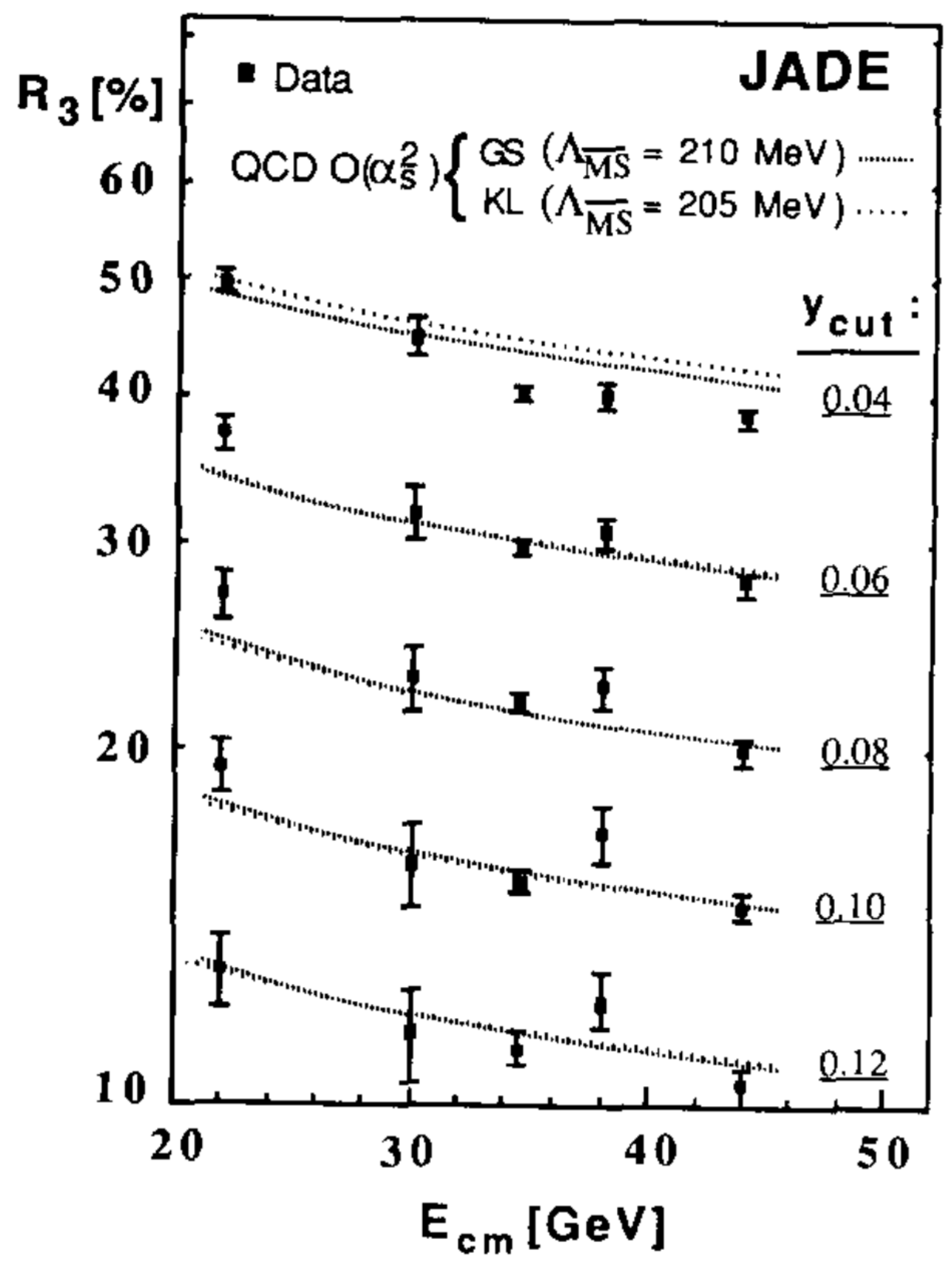}
\end{center}
\caption{Three-jet event rates at different centre of mass energies for various values of the
jet resolution parameter $y_{cut}$, together with the predictions of the complete second-order 
perturbative QCD calculations of Gottschalk and Shatz (GS) and of Kramer and Lampe (KL).}
\label{fig:running-3-jet}       
\end{figure}

Finally, two years after the shut-down of the PETRA collider,
JADE extended its studies of jet production rates
using the full data statistics 
in the energy range from 22 to 46.7 GeV \cite{Bethke1988},
and published the first \oq Experimental Investigation
of the Energy Dependence of the Strong Coupling Strength".
Instead of determining values of $\asq$, 
the energy dependence of the observed 3-jet event production rate $R_3$ was analysed,
because $\as$ determinations suffer from large systematic
uncertainties due to modelling of the hadronisation process and
to treatments and approximations of higher order terms 
used in different QCD calculations.

According to QCD perturbation theory, the 3-jet rate defined by the JADE jet algorithm 
is proportional to the strong coupling. 
In second and higher perturbative order, the prediction reads
$R_3(q) = C_1 \asq + C_2 \as^2(q) + $ \textit{higher order terms}, 
whereby the coefficients $C_n$ are - for a given value
of $y_{cut}$ - independent of the energy scale $q$, and the energy dependence of $R_3$ 
is solely determined by that of the coupling strength $\as$.
Additional  effects, induced by the hadronisation process, are demonstrated to
be small and energy independent, in regions of sufficiently large values of $y_{cut} \ge 0.06$
and for c.m. energies above 22 GeV. 

The measured energy dependence of $R_3$, in the c.m. energy range from 22 to 46 GeV
and for different fixed values of $y_{cut}$, is shown  in Fig.\ref{fig:running-3-jet}. 
Without any further correction or modelling of hadronisation effects, the data are
well described by the predictions of $\oaa$ perturbative QCD  \cite{GS,KL}.
This applies to both the absolute rates at different values of $y_{cut}$ and their
observed energy dependence. 
The significance for discriminating an assumed energy $in$dependence of $R_3$ and thus, 
of constant $\alpha_s$, is at the level of 4 standard deviations.

Due to its applicability in both experiment and theory and its 
relative insensitivity to hadronisation and nonperturbative effects, 
the JADE jet algorithm was adopted by many experiments
at $\epem$ colliders to come, in particular those at TRISTAN 
($\sqrt{s}$ around 60~GeV) of the Japanese KEK laboratory and
at the Large Electron Positron collider LEP ($\sqrt{s} = 89$ to 214 GeV) at CERN.
Later on, in the 1990s and with the experience of the LEP data, 
this class of algorithms was further developed, 
keeping the basic idea of successive clustering of particles 
but modifying the definition of the jet resolution,
thereby enabling the application of refined theoretical methods 
(see e.g. \cite{bkss}).

\subsection{Hadronisation Phenomenology}
\label{sec:hadronisation}
Colour-charged quarks and gluons do not exist as free particles. 
Instead, they convert into colour-neutral hadrons.
This process
cannot be quantified nor calculated by QCD perturbation theory, but for now must be
parametrised by means of phenomenological hadronisation,
or fragmentation, models.

At the time of early PETRA operation, fragmentation of 
fast moving quarks (and antiquarks) was commonly described using the 
Field-Feynman model \cite{field-feynman},
generating particles (mesons) that acquire a fraction $z$
of the longitudinal momentum of the primary quark
and a limited transverse momentum ($p_\perp$) around the initial quark,
statistically parametrised by a phenomenological fragmentation function $f(z)$
and a gaussian $p_\perp$ distribution with a width parameter $\sigma_q$.

The model of Field and Feynman 
was extended to also cover $q\overline{q}g$ final states, where gluons
were first split into a secondary $q\overline{q}$ pair, with subsequent 
Field-Feynman like hadronisation of these secondary quarks
\cite{hoyer}\cite{ali}.
An alternative approach to hadronise $q\overline{q}g$ final states was
introduced by the Lund group \cite{lund2}, where fragmentation does not proceed
along the directions of the outgoing quarks, but instead
along string-like colour flux lines between the primary quark and the gluon, and the primary
antiquark and the gluon.\footnote{In
1981 Bo Andersson approached a member of the JADE collaboration with his idea of the 
\oq LUND-string" and how to test it. 
The studies started without delay and became 
the thesis subject of Rolf Felst's student Alfred Petersen.}

This QCD-inspired treatment of colour-charged quarks and gluons predicts 
a depletion of particles in the region between the primary quark and antiquark,
compared to the regions between the (anti)quark and the gluon.
The "independent" gluon fragmentation models \cite{hoyer,ali} would not
predict such a depletion.

In addition to the string-effect and its QCD-inspired interpretation, QCD predicts the
general properties of gluon-initiated jets to be different from jets initiated by quarks.
Due to the larger colour-charge of gluons and the effects of gluon-self-coupling,
particles from gluon jets are expected to exhibit a broader distribution of transverse
momenta $p_\perp$ w.r.t. the principal (gluon-) jet direction.

\subsubsection{String Fragmentation and Differences between Quark- and Gluon-Jets}
\label{sec:string}
\begin{figure}[htb]
  \includegraphics[width=\textwidth]{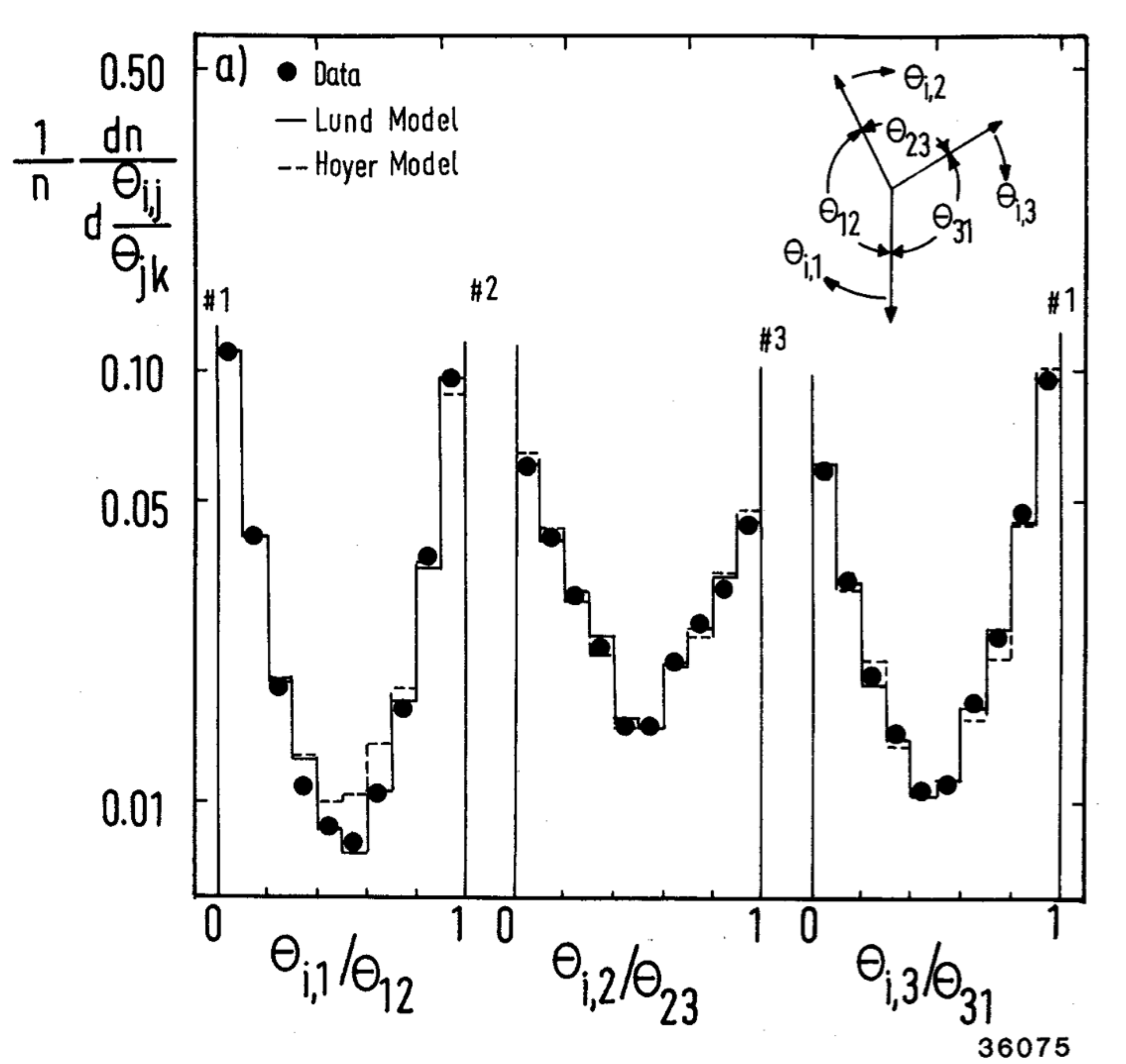}
\caption{The average charged and neutral particle density in the angular regions between the jet axes, 
normalized by the total number of particles, versus $\Theta_{i,j} / \Theta_{jk}$.
The data are shown together with predictions of the Hoyer and the Lund model. 
Jets are ordered according to their energies, $E_1 \ge  E_2 \ge E_3$, so that
jet $\#3$ is the gluon initiated jet, and jets $\#1$ and $\#2$ are the quark and antiquark jets 
most of the time.}
\label{fig:string}       
\end{figure}

The prediction of the \oq string-effect" was demonstrated, for the first time, in  
\oq Experimental study of jets in electron-positron annihilation" \cite{jade-string1}
by JADE, and corroborated later
in a more detailed study of 
\oq Particle Distributions in 3-Jet Events Produced by $\epem$ Annihilation"
\cite{jade-string2}, see Figure~\ref{fig:string}, and in a further
\oq Test of Fragmentation Models by Comparison with Three Jet Events Produced in $\epem \rightarrow$ Hadrons" \cite{jade-string3}.
The string effect appears as a depletion of particles in the region between the two highest
energetic jets of 3-jet events.

Evidence for the string effect was more than just of academic importance:
experimental determinations of the strong coupling $\alpha_s$ - a fundamental parameter of nature
like the electromagnetic coupling and fine structure constant $\alpha$ - resulted in significantly different
values of $\alpha_s$ when data were corrected for hadronisation effects using these fundamentally 
different types of models, see e.g. \cite{cello-model-dependence}. 
The experimental evidence for the string effect was finally confirmed by TASSO \cite{tasso-string}
and by the TPC/2$\gamma$ Collaboration  \cite{tpc-string}
at the PEP collider at the Stanford Linear Accelerator Center.
Furthermore, it was demonstrated - in the theoretical framework of local parton-hadron duality -
that coherence of soft gluon emission provides the QCD explanation of the string effect observed in experiments  \cite{azimov}.
Since that time, the phenomenology of independent jet fragmentation is depreciated and
the concept of string hadronisation has become a standard in modelling hadronic final states at all generations of particle colliders following the days of PETRA.

\begin{figure}[htb]
  \includegraphics[width=\textwidth]{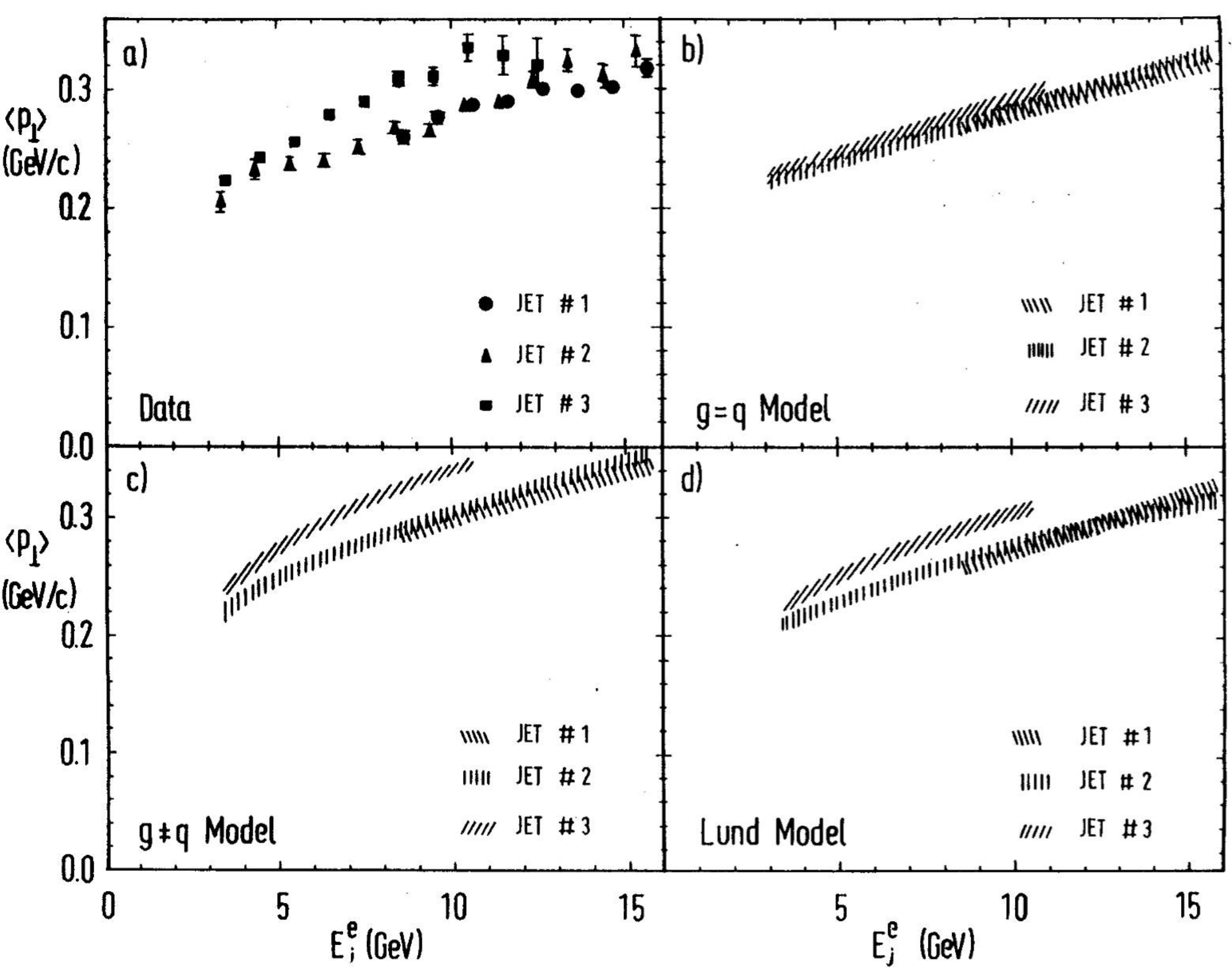}
\caption{The average transverse momentum of the charged and neutral particles 
within a jet relative to the jet axis, for the three jets
as a function of the jet energy for 
(a) the experimental data with $E_{cm} > 29$ GeV, and 
(b), (c), (d) for the prediction of the q = g model, the
q $\neq$ g model ($\sigma_q$ = 330 MeV, $\sigma_g$ = 500 MeV), and the Lund
string fragmentation model, respectively.}
\label{fig:qgdiff}       
\end{figure}

In a further study of 3-jet events, comparing the average
transverse momentum $\langle p_\perp \rangle$ of particles in the lowest
energy jet with those in the other jets, JADE found
\oq Experimental evidence for differences in $\langle p_\perp \rangle$ 
between quark jets and gluon jets" \cite{qgdiff}, see Fig.~\ref{fig:qgdiff}.
At a given jet energy, the average
$p_\perp$ of particles within the
lowest energy jet of a 3-jet event is larger than the average $p_\perp$ of the other jets. 

The string fragmentation model reproduces the observed effect, while
for independent jet fragmentation models, parametrised
according to Field and Feynman, one needs a $\sigma_q$ of
about 300 MeV for quark jets and of about 500 MeV
for gluon jets to explain these data. 
However, even with increased $\sigma_q$ for the gluon jet,
independent jet fragmentation models are not able to reproduce
the string effect as observed in Fig.~\ref{fig:string}.

\subsubsection{Charged Particle, Neutral Kaon and Baryon Production}
\label{sec:particle}

Most analyses of $\epem$ annihilation processes into quarks and gluons 
require a precise knowledge and modelling of the fragmentation process. 
These models contain a number of free parameters and parametric functions,
which are not given by theory, but must be adjusted to provide an optimal 
description of the data.
In these models the multiplicity of charged particles is essentially given by fragmentation functions, 
determining the relative particle momenta along the jet direction, 
and by the fraction of pseudoscalar to vector particles produced. 

In a study \oq Charged Particle and Neutral Kaon Production in $\epem$ Annihilation at PETRA"
\cite{multiplicity},
JADE determined experimental constraints on the parameters 
controlling these model properties, in a previously unknown c.m. energy range
from 12.0 GeV to 36.7 GeV. 

\begin{figure}[htb]
  \includegraphics[width=\textwidth]{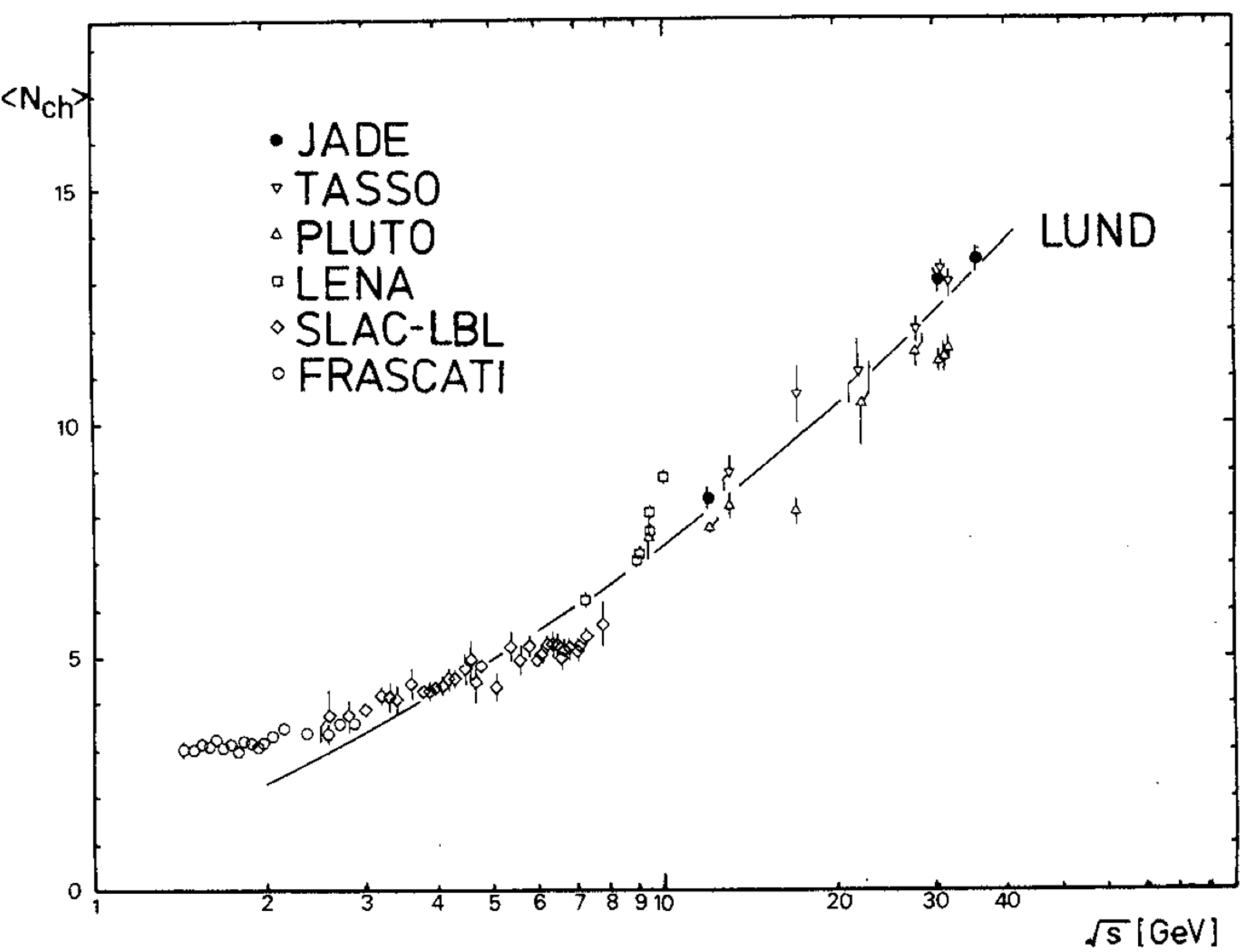}
\caption{Mean charged particle multiplicity as function of the centre-of-mass energy
$\sqrt{s}$. 
Only statistical errors are shown. 
Decay products of neutral kaons are included. 
The curve is obtained using the Lund model.}
\label{fig:multiplicity}       
\end{figure}

As one example, the energy dependence of the mean 
charged particle multiplicity is shown in Fig.~\ref{fig:multiplicity}.
It is found to rise faster than logarithmically with energy.
The data can be
satisfactorily fitted by functions predicted by perturbative QCD,
and also by fragmentation models based on first order QCD with fragmentation
parameters properly adjusted.

Distributions of the charged particle multiplicity at different c.m. energies
exhibit KNO scaling \cite{kno}.
These distributions are well described by the Lund model,
while the independent fragmentation model of Hoyer et al. fails in this respect at
c.m. energies above 30 GeV.
 
The observed number of neutral kaons per event varies from 
1.1 at $\sqrt{s} =$ 14~GeV to 1.5 at $\sqrt{s} =$ 35~GeV
and can be matched by the fragmentation model with a relative
suppression factor $\gamma_s = 0.27 \pm 0.03 \pm 0.05$
of strange quarks being produced in the fragmentation process - 
to be compared to an expected factor of 1 if strange quarks had 
the same mass as the light u and d quarks, and to 0.5 as assumed
in the original Field-Feynman model.

The production of baryons, bound states of three quarks or three antiquarks,
was not modelled in early versions of fragmentation models, and details about 
baryon production were scarce in the early times of PETRA operation.
In a study of \oq Baryon Production in $\epem$ Annihilation at PETRA"
\cite{baryons}, JADE selected candidates of antiprotons $\overline{p}$ 
in the momentum range from 0.3 to 0.9 GeV/c,
through their specific energy loss (dE/dx) measured in the central jet chamber.
In addition, candidates for antilambdas ($\overline{\rm \Lambda}$) were selected
through their decay $\overline{\rm \Lambda} \rightarrow \overline{p} \pi^+$.
About 20\% of all hadronic events contained a $\overline{p}$, either directly produced 
in the fragmentation process or from hyperon or resonance decays, and 7\% had a 
$\overline{\rm \Lambda}$.
The data indicated an angular anticorrelation of baryon-antibaryon pairs.

\subsection{Electroweak Precision Tests}
\label{sec:electroweak}

\subsubsection{Total Cross Section}
\label{sec:total-x}
 
The ratio of the hadron production cross section 
to the lowest order pointlike QED cross section,
$R = \sigma_{had} /  \sigma_{pt}$, with
$\sigma_{pt} = (4\pi /3 s) \alpha^2$,
 is a fundamental quantity in $\epem$ interactions, as it provides important information about the number of quark flavours which are produced in the accessible energy range. 
$R$ is calculated in the quark-parton model as 
$R = 3 \Sigma_q Q_q^2$, where $Q_q$ is the quark electric charge, and the summation runs over all the produced quark flavours $q$. 
If the threshold for pair production of new charge 2/3 (1/3) quarks is passed, $R$ 
is increased by about 36\% (9\%). 
The importance of the $R$ measurement was enhanced by theoretical considerations that 
predicted the existence of a  yet undetected, sixth quark,
i.e. the top quark which was assumed to become accessible at PETRA.
Also, if quarks would have structure, a deviation from a constant $R$ should be observed. 

The value for $R$ is modified when including the lowest order 
QCD corrections and electro-weak effects: 
QCD corrections to first order increase $R$ by about 5\% at 
$\sqrt{s} = 30$~GeV. 
The effect of the weak neutral current is energy dependent. 
It increases $R$ by 1.5\% at $\sqrt{s} = 37$~GeV.

Therefore, 
from early on all PETRA experiments focused on the measurement 
of the total cross section. 
During the lifetime of the experiment, 
JADE issued three publications covering this topic \cite{jade-total-x,jade-r2,jade-r3}.

For these measurements, an important feature of the JADE detector was its sensitivity to charged 
particles and photons over 97\% and 90\% of the full solid angle, respectively. 
In a first step, multihadron ($\epem \rightarrow$ hadrons)
events were identified by a \oq charged particles trigger"
in combination with a \oq shower energy trigger". 
Both triggers showed a considerable overlap. 

In a second step, multihadron events were selected offline by a set of criteria which eliminated various 
sources of background, such as $\epem \rightarrow \epem$ events 
(Bhabha scattering), $\tau$ pair production or cosmic ray background, 
by applying cuts on the momentum balance as well as on the visible energy of events. 
The luminosity was determined from small angle Bhabha scattering detected by the end cap counters.

\begin{figure}[htb]
\begin{center}
  \includegraphics[width=0.8\textwidth]{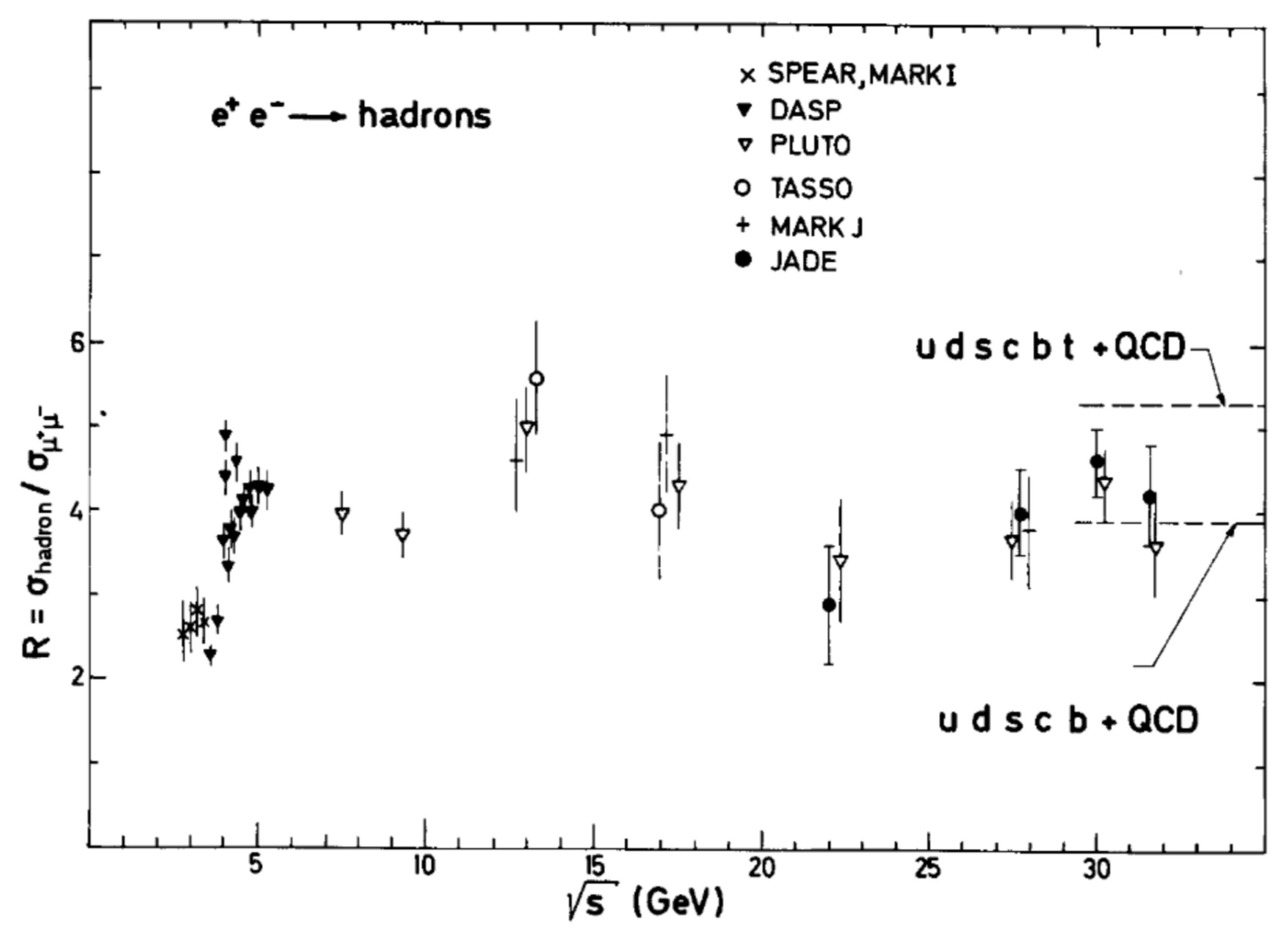}
\end{center}
\caption{Ratio $R$ of the hadronic to the point like cross section as a function of 
the centre-of-mass energy $\sqrt{s}$. 
The predictions with and without a top quark are also shown. 
}
\label{fig:r1}       
\end{figure}

The values for $R$, obtained from the data collected during the first year of PETRA operation, 
a total of 357 multihadron events, are shown in Fig~\ref{fig:r1}, 
together with data from other experiments. 
The errors are statistical only. 
Assuming only u, d, s, c and b quarks, the naive quark-parton model predicts $R = 11/3$. 
This value is increased from 3.66 to 3.9 by the inclusion of QCD corrections.  
The measured values of $R$ are compared with the expected $R$ 
in case of the production of a charge 2/3 top-quark pair. 
There is no evidence in the data for the production of a new quark flavour with a charge of 2/3. 

In a second paper \cite{jade-r2}, 
$R$ values were measured in the center of mass energy range between 12.0 and 36.4~GeV 
with systematic errors of typically $\pm 3$\%. 
This analysis was based on an integrated luminosity of $38 pb^{-1}$ 
accumulated in 1979-1981. 
About 15 000 multihadron events were obtained. 

The number of measured multihadron events was corrected for acceptance, calculated by a Monte Carlo simulation. 
In this simulation, the Lund model \cite{lund} was used together with the initial state radiative 
corrections of Berends and Kleiss \cite{bk}. 
The parameters in the model were chosen such that the measured and calculated charge multiplicity 
and neutral kaon production agreed. 
The simulation also showed that the fraction of events lost by the hardware trigger 
was less than 1\% with a negligibly small error.

The data were consistent with a constant $R$ in this energy range with an average value 
of $3.97 \pm 0.05$ (statistical and point-to-point systematic error) 
$\pm 0.10$ (overall normalisation error). 
Corrections due to QED process of 
${\cal O}(\alpha^4)$ or higher were not included. 
The data excluded a step in $R$ of 
$\Delta R_0 > 0.29$ at the 95\% confidence level, 
ruling out the pair production of
a charge 2/3 quark with mass between 7.5 and 17.5 GeV. 

\begin{figure}[htb]
\begin{center}
  \includegraphics[width=0.8\textwidth]{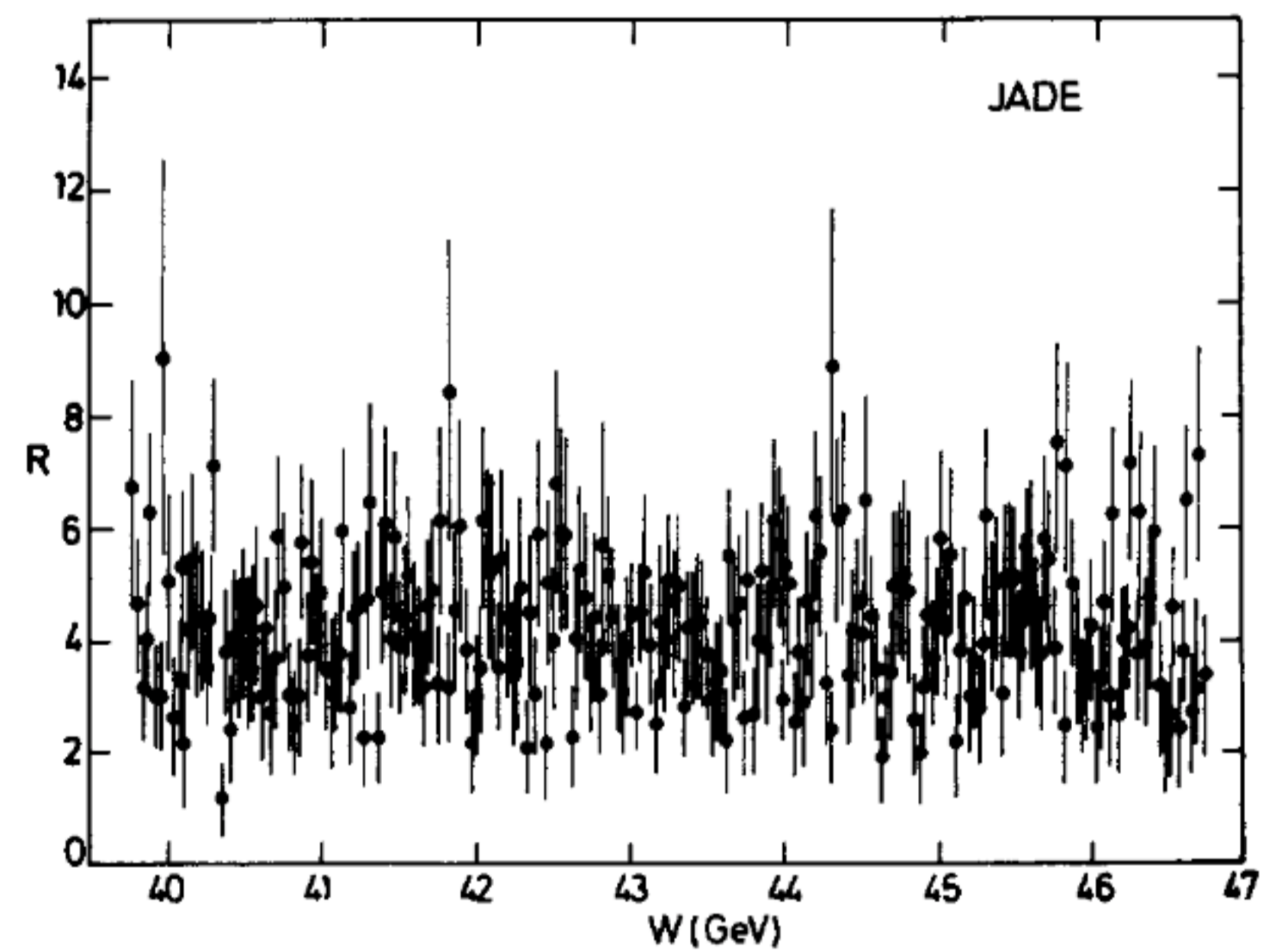}
\end{center}
\caption{The ratio $R$ measured between $\sqrt{s}= 39.79$~GeV and 46.78 GeV 
during a scan in steps of 30~MeV in order to find narrow resonances of bound $t \overline{t}$
states .}
\label{fig:r2}       
\end{figure}
\begin{figure}[htb]
\begin{center}
  \includegraphics[width=0.8\textwidth]{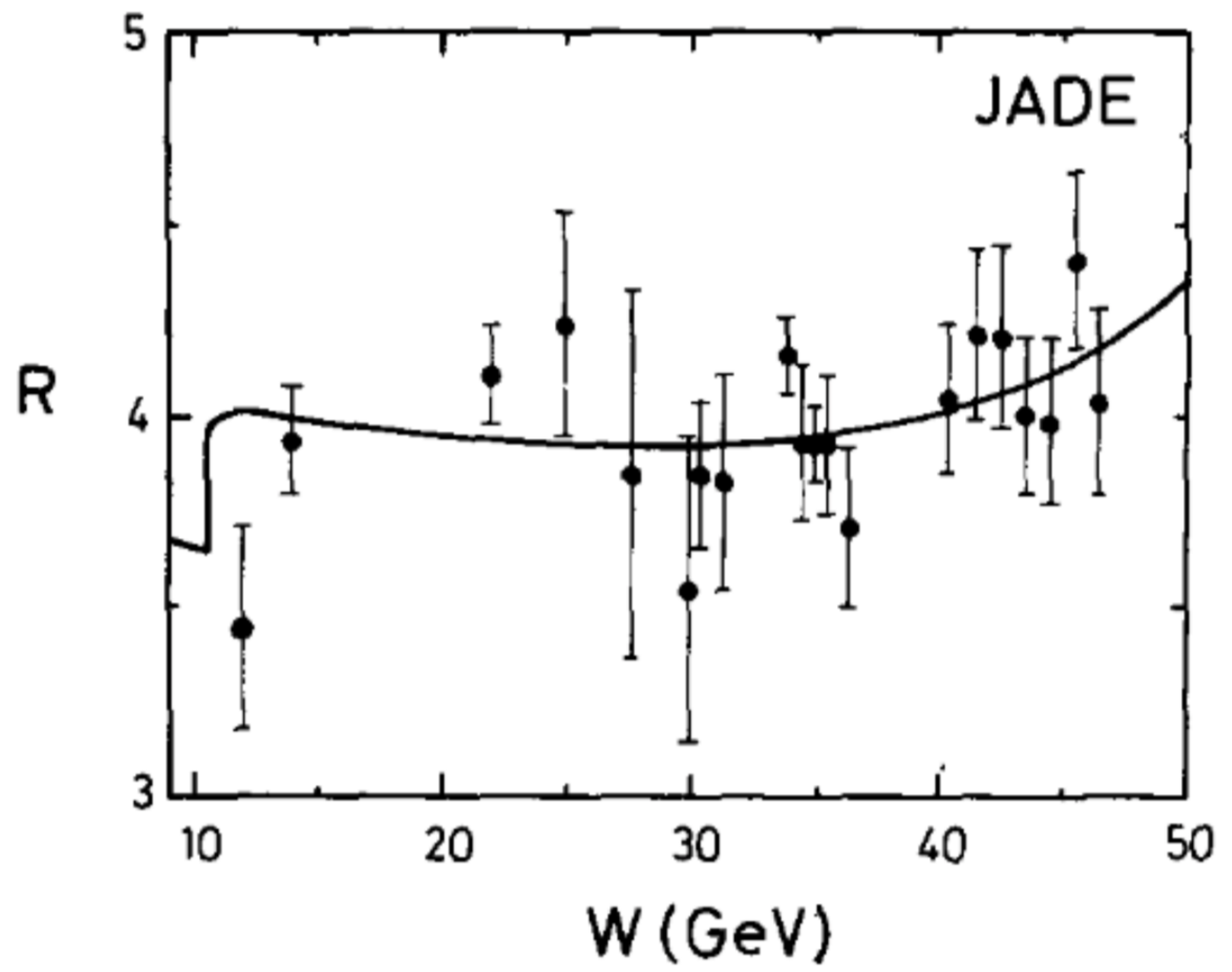}
\end{center}
\caption{Summary of the ratio $R$ averaged in bins of 1 GeV c.m. energy. 
The curve represents the best fit to the standard electroweak interaction
model with QCD corrections, with $\as (30\ GeV) = 0.20$ and
$\sin^2\theta_w = 0.23$.
}
\label{fig:r3}       
\end{figure}

The third publication \cite{jade-r3} reported the results of the total cross section measurements 
in the highest energy range explored by PETRA, between 39.79 and 46.78~GeV,
and summarised the results obtained in the entire PETRA energy range. 
The specific goal of the high energy run was the search for the top quark by a step-wise 
increase of the energy. 
Toponium - the lowest mass bound state of a top- and an antitop-quark -
would locally enhance the total cross section by a factor of 2 to 3.

The R-values of this scan are shown in Fig.~\ref{fig:r2} as a function of the c.m. energy. 
The overall normalisation error is 3.3\%, the point-to-point error $\pm 1\%$. 
The data are consistent with a constant value of $R$ in the scanned energy range. 
The average value is $R = 4.13 \pm 0.08 \pm 0.14$, where the second error 
is the overall normalisation error.

Looking at the entire data sample collected by JADE in the time between 1979 and 1986 
in the energy range from 12 to 46.8~GeV one obtains Fig. \ref{fig:r3}. 
The plot clearly exhibits two effects, the absence of the top quark (later found at a much higher mass 
of 175~GeV), and the good description of data by a fit to the standard electroweak interaction 
model, yielding 
$\sin^2 \theta_W = 0.23^{+ 0.03}_{- 0.04}$ for the weak mixing angle $\theta_W$,
and exhibiting the tail of the rising cross
section of real $Z^0$-boson production\footnote{The $Z^0$ boson was
discovered in 1983 at CERN's Super Proton Synchrotron, at a mass of $\sim$92 GeV/c$^2$
and a decay width of $\sim$2.5~GeV/c$^2$
\cite{ua1-z0,ua2-z0}.}.
\subsubsection{Charge asymmetry in $\epem \rightarrow \mu^+ \mu^-$}
\label{sec:mu-asymmetry}

At the highest PETRA energies the standard electro-weak theory predicts sizeable interference effects 
between electromagnetic and weak neutral currents which 
manifest themselves in angular asymmetries. 
At low energies, such processes are 
well described by QED alone, if contributions up to 
${\cal O}(\alpha^3 )$ are included. 

In a first publication  \cite{mu-asym},
the angular distribution and the $s$ dependence of the total cross section
for the process  $\epem \rightarrow \mu^+ \mu^-$
was measured by JADE at centre of mass energies in the range 
$12.0 \le \sqrt{s} \le 36.8$~GeV. 
The measurement was made using the jet chamber to measure directions, momenta and charges of the outgoing particles. 
Separation of muon pair candidates from background processes was achieved 
with the help of time-of-flight (TOF) counters, lead glass shower counters, and the muon filter.

The angular distribution of the positive muon for the data above 
a c.m. energy of 25 GeV showed, after extrapolation to the full solid angle,
a clear forward-backward asymmetry of 
$A = - (11.8 \pm  3.8 \pm 1) \%$, 
where the first error is statistical and the second one systematic. 
This asymmetry demonstrates sizeable interference effects between electromagnetic and weak neutral currents, thereby excluding pure QED by three standard deviations while agreeing well with the standard model SU(2) x U(1) of Glashow, Salam and Weinberg, 
which predicts an asymmetry of $A = -7.8 \%$ for the same angular acceptance. 

\begin{figure}[htb]
\begin{center}
  \includegraphics[width=0.9\textwidth]{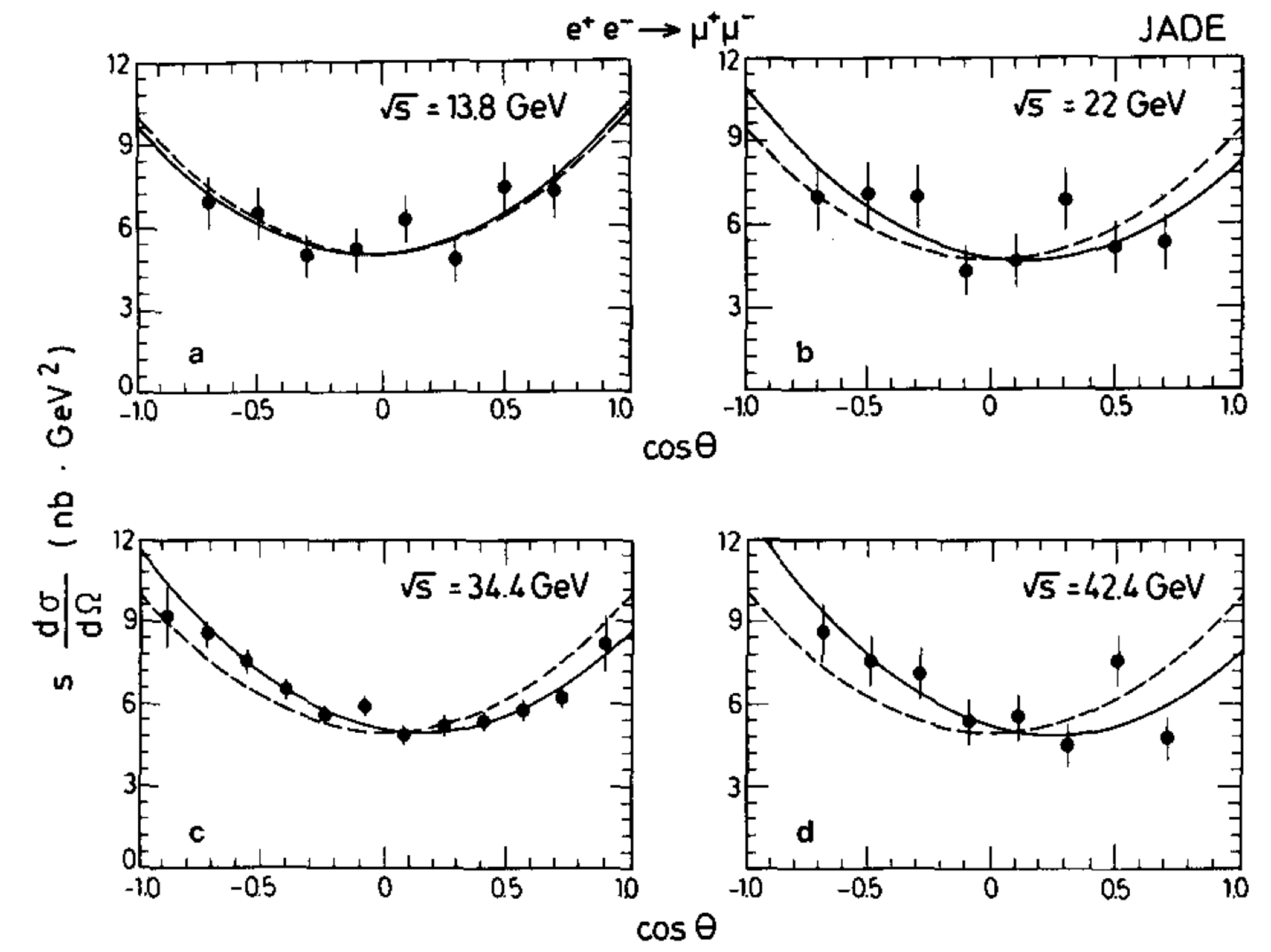}
\end{center}
\caption{Angular distributions of $\epem \rightarrow \mu^+ \mu^-$
for four c.m. energies. 
The dashed lines are symmetric fits $f(\theta) \propto (1 + \cos^2 \theta)$, the full lines are
fits allowing an additional asymmetry $f(\theta) \propto (1 + \cos^2 \theta + B \cos \theta)$
\cite{mu-asym2}.}
\label{fig:mu-asym2}       
\end{figure}

An update of these measurements based over a larger energy range, up to 42 GeV and on higher 
statistics \cite{mu-asym2},
showed how the forward-backward asymmetry increased with the center of mass energy, in full agreement with the prediction of the Standard Model, see Fig.~\ref{fig:mu-asym2}.

\subsubsection{Charge asymmetry in $\epem \rightarrow b\overline{b}$}
\label{sec:b-asymmetry}

Electroweak interference effects in $\epem$ collisions in the energy range of PETRA 
are expected to also lead
to a forward backward asymmetry in the emission of the final state quark pairs. 
This asymmetry can be used to measure the electroweak charges. 
The first statistically significant asymmetry measurements had been performed for purely leptonic final states for two main reasons: 
(a) Although the interference effects for quarks are expected to be large, the separation of the various flavours has been a problem. 
(b) The determination of the primary quark charge is challenging.\footnote{
Due to the high quality tracking, in rare cases the full decay of the b-hadrons could be reconstructed. 
Dieter Haidt, senior scientist at DESY came to Hamburg after having worked on the discovery 
of neutral currents at the Gargamelle bubble chamber at CERN. He recalls: 
\textit{My choice to join 
JADE rather than one of the other PETRA groups was based on the fact that the jet chamber 
was nearest to a bubble chamber and my experience may perhaps be to the benefit of the collaboration: 
Once I found an outstanding event and presented it to the JADE meeting as a complete cascade 
decay of a B-hadron, i.e. b $\rightarrow$ c with subsequent c $\rightarrow$ s and finally manifesting itself as 
a K in the final state. Rather than applause I got the concise remark by Orito: \oq Come back, 
if you have a thousand of such events".}  This story is typical for the rigorous quality 
requirements in JADE, and also demonstrates the change of paradigm from optical to
electronic recording and analysis of particle reactions.}

The JADE paper \oq A Measurement of the Electroweak Induced Charge Asymmetry in 
$\epem \rightarrow b\overline{b}$" \cite{jade-b-asym}
describes a method for partially overcoming these problems for b-quarks, thus enabling the measurement of the forward-backward charge asymmetry for the process 
$\epem \rightarrow b \overline{b} \rightarrow \mu^\pm +$~hadrons.

The flavour separation is based 
on three observables derived from the 
kinematics of selected hadronic events which contain an identified inclusive muon:
1) the jet transverse mass, 
2) the muon transverse momentum, and 
3) the overall transverse momentum imbalance, i.e. the missing transverse momentum. 
In all three cases, the momentum components are measured transverse to the major axis of the event 
sphericity ellipsoid (see e.g. \cite{jade-top-search}) which defines the jet axis. 
The transverse mass is obtained by summing the transverse momentum components of all charged and neutral particles (except the muon).

Monte Carlo simulations showed that the probability distribution functions 
of these quantities exhibit a good discrimination 
between light quarks (uds) and the heavier quarks, c and b \cite{marshall-b}. 
The simulation also reproduced the detailed shapes of the measured spectra of the
observables, indicating that they
are well understood and
the JADE detector is sufficiently hermetic to allow these variables to be used successfully.

The data used for this analysis were collected up to the summer of 1983 and correspond to an 
integrated luminosity of 76~pb$^{-1}$. 
From the sample of multihadron events with PETRA beam energies above 15 GeV, 
1780 events were selected containing a muon candidate together with hadrons which 
satisfied the standard JADE visible energy and longitudinal momentum balance criteria. 
The average c.m. energy of this sample was 34.6 GeV.

\begin{figure}[htb]
\begin{center}
  \includegraphics[width=0.7\textwidth]{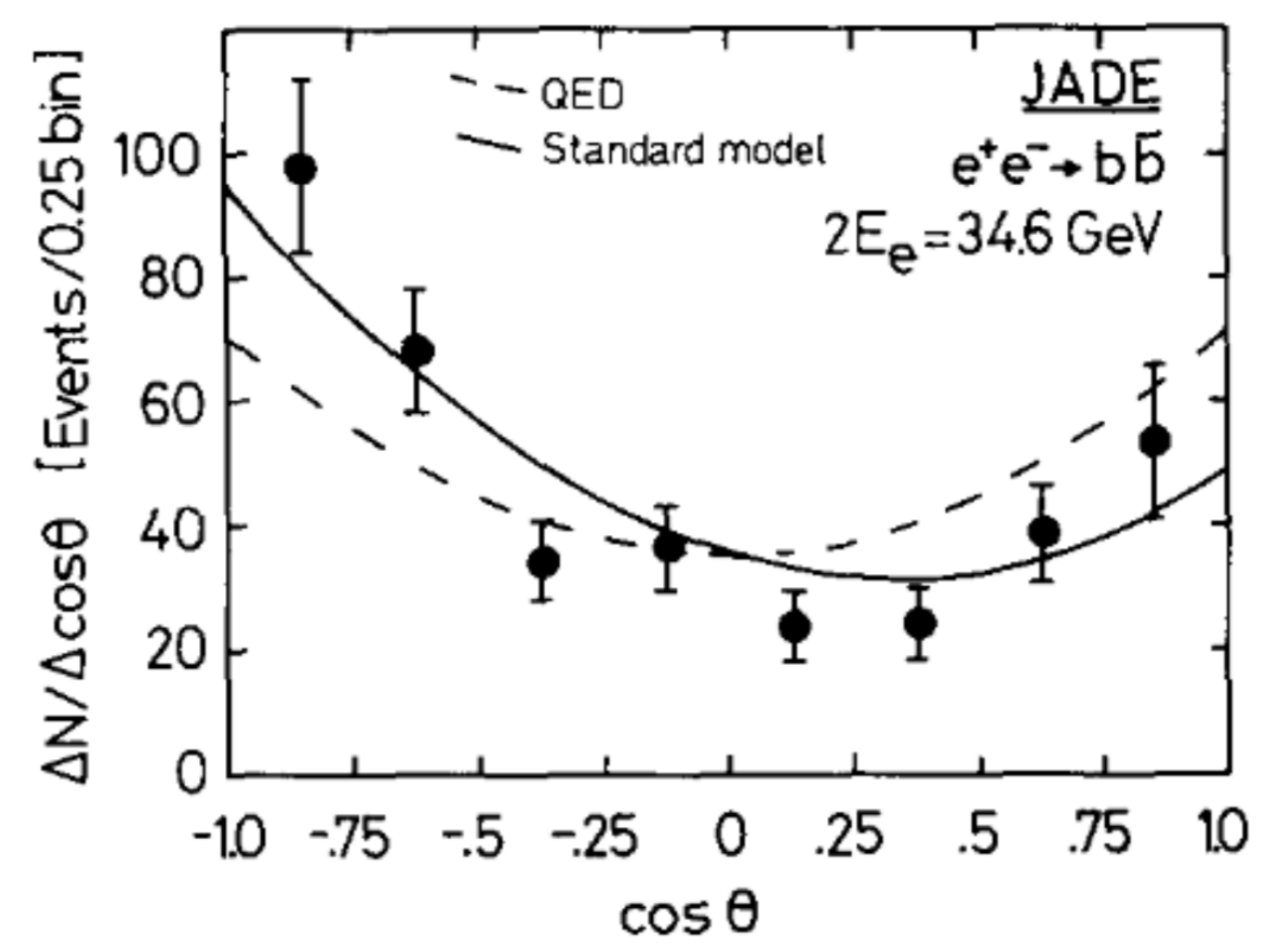}
\end{center}
\caption{The extracted angular distribution for $b\overline{b}$ events. 
The dashed curve is proportional to $1 + \cos^2\theta$ (no asymmetry) and
the solid curve is the prediction of the standard model at 34.6~GeV.}
\label{fig:b-asym}       
\end{figure}

The polar angle $\theta$ of the final state $q \overline{q}$ system was estimated 
using the sphericity axis. 
If the event had a $\mu^-$ in the direction of the $e^-$ or a $\mu^+$ in the direction 
of the $e^+$ beam, then $\cos \theta$ was defined to be positive. 
Using the three variables described above, a likelihood analysis was used to determine the numbers for forward and backward b events:
$N_b^F = 114.6 \pm 12.5$,\ \ \ $N_b^B = 191.3 \pm 16.2$.
 The angular distribution of these events is shown in Fig.~\ref{fig:b-asym}, exhibiting a clear forward-backward asymmetry.

The ratio of forward-backward events yields, after correcting for acceptance, an asymmetry for b-quarks of $A_b = (-22.8 \pm 6.0 \pm 2.5)\%$. 
The Standard Model prediction for the b asymmetry is $- 25.2$\% . 
This number is only 2.8 times that of the $\mu^+\mu^-$ asymmetry (and not 3 times
as expected from the ratio of charges) due to the significant mass of the b quark. 
The measurement is in good agreement with this prediction.

This measured asymmetry can be used to determine a value for the ratio of the b-quark's 
weak axial ($a_b$) 
to the electric ($Q_b$) charge of 
$a_b / Q_b = - 2.71 + 0.71$~(stat) + 0.30 (syst). 
Assuming $a_b = a_c = - 1$ , the result determines the b electric charge:
$Q_b = -0.37_{-0.07}^{+0.13}$~(stat.) $\pm 0.03$~(syst.), 
which is consistent with charge -1/3 for the b quark.

\subsection{Photon Structure Function}
\label{sec:f2}
Since the early 1970s, deep inelastic scattering processes of highly virtual photons
$\gamma^*$ (with $Q^2 >> p^2$)  on a target real photon ($p^2 \approx 0$), $Q^2$ being the 4-momentum squared of the virtual and $p^2$ the 4-momentum squared of the real photon, 
raised significant interest because the structure function of the photon,
$F_2^\gamma (x,Q^2)$, was predicted to exhibit features that are quite different
from those of the structure function of protons, $F_2^p (x,Q^2)$ - see e.g.
\cite{buras} for a summary of the development of the field.

In short, such processes can be studied in reactions like $\epem \rightarrow \epem +$ hadrons,
where photons radiated off the initial electron and positron transform into a final state of hadrons.
Cross sections of such processes at $\epem$ colliders
are expressed in terms of  photon structure functions
$F_2^\gamma (x,Q^2)$ and $F_l^\gamma (x)$, where
$Q^2 = 4 E E' \sin^2(\Theta /2)$ is the squared 4-momentum transfer,
$x = Q^2 / (Q^2 + W^2)$, $E$ and $E'$ are the initial and final energies of the tagged 
highly virtual photon, $\Theta$ its polar scattering angle,
and $W$ is the invariant mass of the produced hadronic system.

At low $Q^2$, i.e. a few GeV$^2$, the photon was commonly described to behave like a vector meson,
due to quantum fluctuations of the photon into a meson with the same quantum numbers. 
In that case, $F_2^\gamma$ should show $Q^2$ and $x$ dependencies
similar to $F_2^p$, i.e. $F_2$ would decrease, at fixed $x$, with increasing $Q^2$ -
a feature known as scaling violations.

It was found, however, that at not too small $Q^2$, the process 
of $\epem \rightarrow \epem +$ hadrons
is dominated by the point-like cross section according to the quark-parton model,
where the photon fluctuates into a quark-antiquark pair,
and $F_2^\gamma$ is predicted to increase with $\log (Q^2)$ \cite{walsh}. 
Furthermore, the $x$-dependence of
$F_2^\gamma$ can be calculated within the quark-parton model - 
which is also in contrast to $F_2^p$, where the $x$-dependence 
cannot be predicted but must be determined by experiment.
Going beyond the simple quark-parton model and including effects of gluon
dynamics, and thus of QCD, the slope of the $Q^2$-dependence can be calculated and is predicted 
to depend only on a single scale parameter $\Lambda$ or, alternatively, the QCD coupling strength
$\as$.

\begin{figure}[htb]
\begin{center}
  \includegraphics[width=0.6\textwidth]{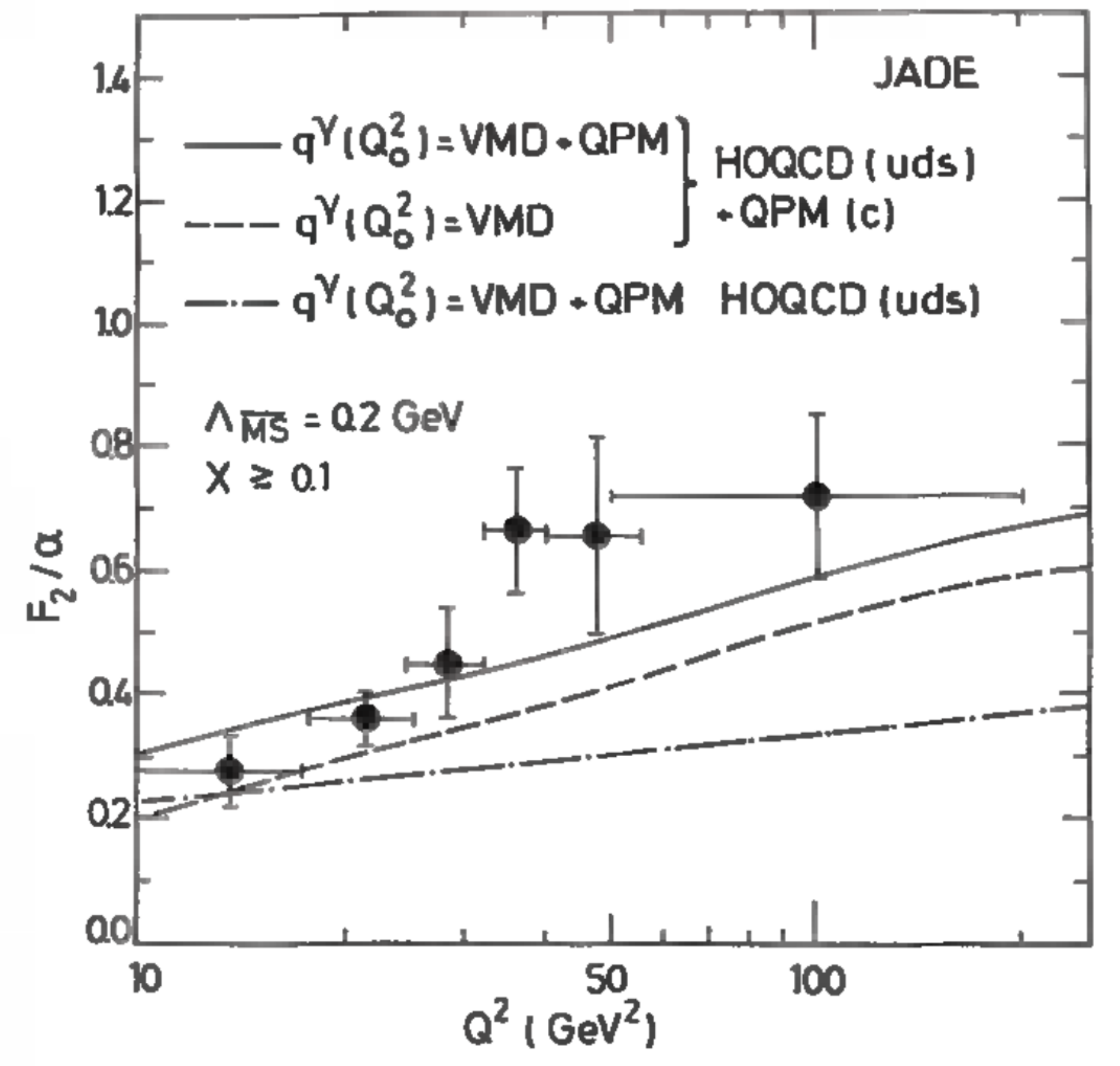}
\end{center}
\caption{ 
Measured photon structure function
$F_2 (Q^2)$ obtained by averaging $F_2(x, Q^2)$ over 
$x \ge 0.1$. 
Solid and dashed curves: non-asymptotic higher order QCD (HOQCD) for light quarks (u, d, s) + 
quark-parton model (QPM) for charm quarks (c), for two variants of parton distributions $q^\gamma$;
dot-dashed curve: non-asymptotic HOQCD for light quarks only \cite{jade-f2}.}
\label{fig:f2}       
\end{figure}

JADE measured and analysed hadron production initiated by two photon 
interactions, $\epem \rightarrow \epem +$~hadrons,
where one of the scattered electrons was detected at large angles, 
with $Q^2$ ranging between 10 and 220 GeV$^2$  \cite{jade-f2}. 
The data were compared with  non-asymptotic as well as asymptotic QCD predictions.  
The former takes into account both the contributions from the point-like 
and vector-meson-like hadronic components of the photon whereas the latter considers 
only the point-like contribution.

Charged particles were measured with the central drift chamber, 
photons and electrons were measured in the lead glass shower counters. 
The tagging of the electrons was carried out using either the two arrays of endcap lead glass counters 
which covered the polar angular range 245-500 mrad, or the barrel array of lead glass counters which 
covered the polar angular range above 609 mrad.

For each hadronic event the Normalised Longitudinal Momentum Balance (NLMB) was calculated. 
A cut in this variable provides a very efficient suppression of hadronic events 
from $\epem$ annihilation while hardly affecting hadronic two photon events.
The $F_2^\gamma (x)$ functions were determined by unfolding the observed data for detection losses and finite resolution in the JADE detector at values of 24 and 100 GeV$^2$. 
They were found to be well described by asymptotic leading order QCD including only the pointlike contribution of $F_2^\gamma$, which is calculable in perturbative QCD,
and with other leading and higher order QCD calculations and models. 
For asymptotic leading order QCD, $\Lambda_{QCD} = 0.15 (+0.07, -0.04)$ GeV was obtained.

The measurements also show that  $F_2^\gamma$, 
averaged over the range $x >  0.1$, increases as a function of $Q^2$, see Fig.~\ref{fig:f2}.
The rise of $F_2^\gamma$ with $Q^2$ is consistent with the combined effect of the $\ln (Q^2)$ 
dependence of $F_2^\gamma$) and the $Q^2$ dependence of the charm quark contribution,
although the data show a slightly steeper slope than the expectation. 

%% file: Sec-04.tex
\section{JADE Data preservation and Software Revival}
\label{sec:4}
\subsection{Motivation and overview}
\label{sec:motivation}
Long term preservation of data from large scale high-energy
physics (HEP) experiments is imperative to preserve the
ability of addressing scientific 
questions at times long after the completion of those experiments.
Very often, these data are unique
achievements in many scientific respects like energy range,
process dynamics and experimental techniques.
New, improved and refined scientific questions may require
(re-)analysis of such data sets. 
Investments necessary to
repeat past experiments would exceed the efforts of data preservation
by far.

The main reasons driving the quest 
for data preservation are (see e.g. \cite{sb-preservation}):

\begin{itemize}
\item long term completion of the scientific program, 
\item data re-use for new and advanced studies,
\item training, education and outreach.
\end{itemize}

The scientific motivation for re-using and re-analysing data from past 
experiments is given by:

\begin{itemize}
\item the availability of new theoretical input in terms of increased precision, advanced
     models or new predictions;
\item new and improved analysis techniques;
\item the desire to perform cross-checks between different experiments. 
\end{itemize}

After the shut-down of the PETRA in 1986, 
collaborative analyses of JADE data and publication of results 
continued with decreasing pace and came to an end by 1990/1991.
The data were first moved to a few thousand archive tapes,
later to a storage place outside the computing centre, and
finally disposed of when the IBM main frame computer was phased out in 1997.
The source code of the JADE software framework was collected and stored on
private computer accounts which were maintained on the IBM main frame
until 1997. 

The JADE collaboration had no 
plan for further data preservation and future use of their data.
Private initiatives for long-term preservation
started in 1995/1996 at DESY, when Jan Olsson at DESY
organised to copy the JADE data to modern and more efficient data carriers,
and to preserve the JADE software libraries \cite{olsson-2009}.

Driven by the desire to re-analyse JADE data in terms of much advanced QCD calculations and 
Monte Carlo models,
the resurrection of the JADE software and its usability on modern computer platforms
was initiated in 1996/1997,
by one of the authors (SB), then Professor at RWTH Aachen, and put into practice 
by Pedro Movilla Fernandez, a diploma- and later PhD-student at Aachen, 
and by Jan Olsson.
This first step of software recovery was completed by 1999.
Until 2013,  the revived software, running on IBM AIX systems,
was actively used for new analyses of the JADE data, resulting in 3 further
PhD theses, 10 journal publications and several contributions to international conferences and
workshops. 

A second and so far last phase of data preservation started in 2015 and has basically 
been executed 
by Andrii Verbytskyi  at the Max-Planck-Institute at Munich.
In this  process, the JADE software was migrated to LINUX systems and modern build 
tools\footnote{Build tools are programs that automate the creation 
of executable applications from source code.}.

In March 2022, the members of the JADE collaboration 
unanimously decided to release all JADE data and software to be publicly accessible
as \oq open data" and maintained within the CERN open data initiative \cite{opendata}.
The implementation of JADE data, software and documentation into this environment 
is currently in progress.

Digitisation and preservation of documents, auxiliary data and material like publications, 
technical and internal notes, collaboration meeting protocols, 
the log-books from the experiment's main control room, photographs from the time 
of constructing and operating the detector, and manuals describing the functionality of the 
resurrected software and data,
are an equally important heritage 
to be preserved.
Digitised versions of all JADE publications are accessible
through public servers like the InSpire HEP data base \url{https://inspirehep.net}.
Digital copies of  most of the other documents mentioned above were collected and are publicly
available at the JADE web pages maintained at the Max-Planck-Institute of Physics at Munich,
\url{https://www.mpp.mpg.de/en/research/data-preservation/jade}.
Some of the information, like a list and copies of PhD theses performed with JADE data, 
still need to be assembled and will be provided in due time.

\subsection{Data Preservation}
\label{datapreservation}
At the end of data taking,
the JADE data comprised about 1~TB of raw and reconstructed 
data\footnote{About 600 GB of raw data
(\oq REFORM"),  335 GB of reduced and reconstructed data (\oq REDUC1" and \oq REDUC2"),
and 85 GB of specialised physics selection data.} 
- an amount that looks small by standards of the 2020s, but was huge in the 1980s.
The data were stored on thousands of IBM tapes with 160~MB capacity, plus similar amounts
of Monte Carlo generated data and private data selections. 

Space was a problem at the DESY Computer Centre. 
Data stored on \oq Machine Room (M) Tapes" were consecutively moved to \oq Archive (A) Tapes",
stored outside of the main computer area,
if they weren't used for a while. 
Data on A-Tapes were deleted if they stayed unused for a certain time, and if there was no response
on warning messages issued to the responsible user.
This line of action was also applied to software files and libraries - unless they
were declared to be \oq holy" with plausible justification.
At this point, by 1990,
the JADE data and core software were still secured on A-tapes, while most of the Monte-Carlo generated
data had disappeared.

In winter 1991/92, 
the imminent start-up of the HERA collider required to recover large amounts of physical space and infrastructure for proper handling and storage of the new data to come.
The DESY computer centre requested from the PETRA experiments 
to significantly reduce their stores of A-Tapes.
The remaining tapes were packed in big aluminum 
boxes that could only be moved by fork-lifts, 
and stored elsewhere at DESY. 
At this point, the 1 TB of JADE data resided in 23 of these big boxes
that were stored away in DESY's Hall~2, and were deleted from the general catalogue.
\begin{figure}[htb]
  \includegraphics[width=\textwidth]{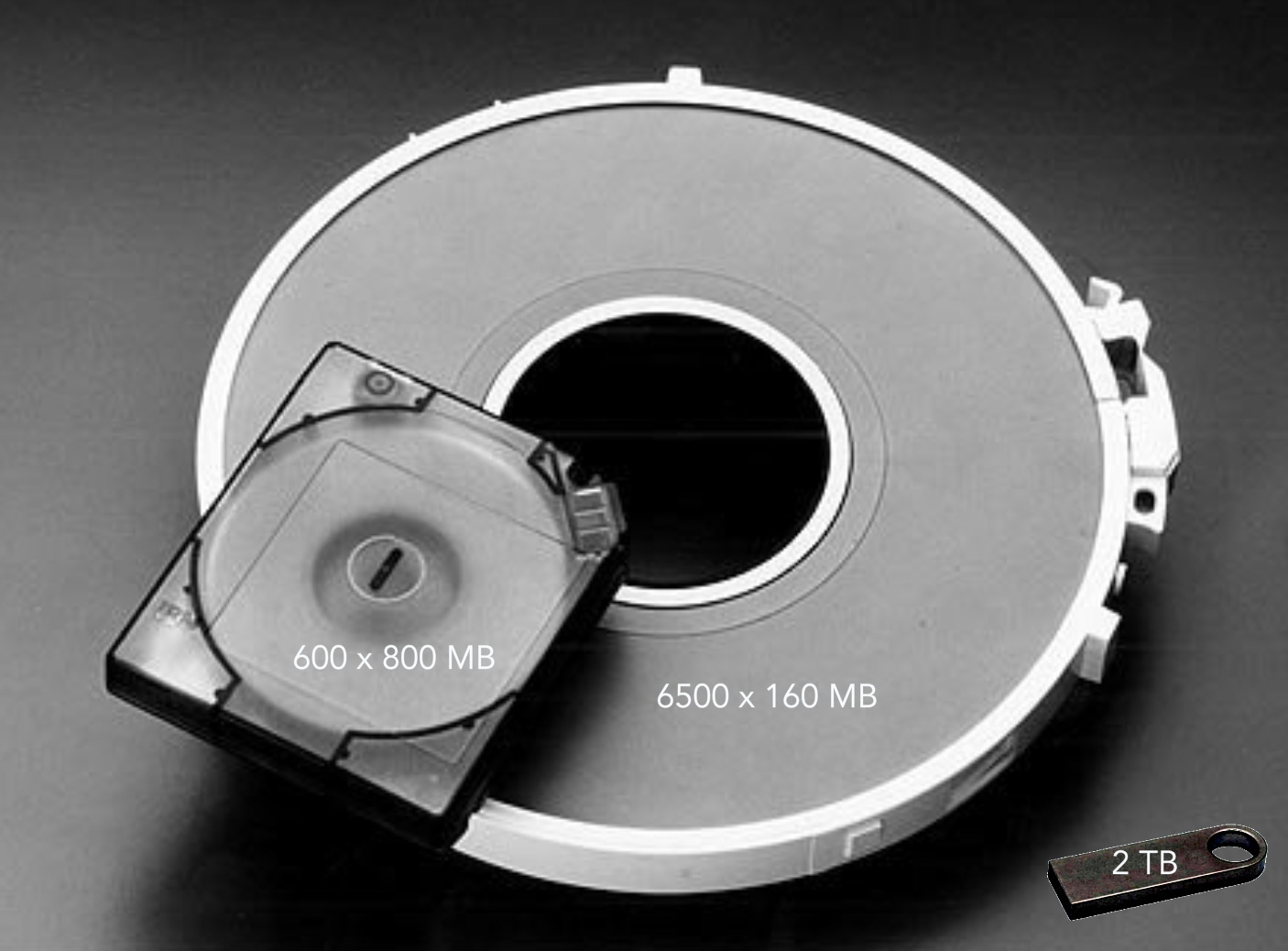}
\caption{The JADE data were originally stored on about 6500
IBM tapes, later converted and written to about
600 IBM3490 cartridges, and nowadays conveniently fit onto a single 2 TB USB memory stick.}
\label{fig:tapes}       
\end{figure}

By 1995, also the space in Hall 2 was needed for other purposes.
Again the PETRA experiments were asked to discard their remaining data,
or else to arrange for further storage by themselves. 
In spring 1996, DESY decided to phase out the IBM Mainframe,
also implying that the old IBM tapes would soon be history.
The responsible JADE experts, still being on-site but involved in other projects like HERA,
decided to move all JADE data onto IBM3490 cartridges, thereby reducing the physical volume of
the required storage space such that it finally \oq should fit into a drawer".

Moving the JADE data to modern data carriers also required to rewrite them such that they
could be read on any future computer platform.
This requirement led to a number of difficulties and problems, some appearing as 
major obstacles at that time, 
which had to be overcome and solved. 
The stories connected to these activities may carry educational messages for future 
ventures of that type:

JADE used an early version of the Bank Object System  
data format \cite{bos}, BOS4. 
As computer memory and storage was precious in the late 1970s and early 1980s, 
a rather intricate structure including combinations of I*2, I*4, F, and A-words in the banks was used,
even assigning single bits and bytes for various purposes and flags - a fatality 
if the data are to be processed on machines which use a different byte order than that
of the old IBM370.
In order to assure platform independence of the data,
they were converted using FPACK \cite{fpack}.
FPACK, however, only worked with BOS77, and the BOS4 source  libraries - 
although declared \oq holy" - no longer existed at DESY, or anywhere else.
This initiated a number of archeological initiatives and actions which finally 
retrieved the original BOS4 source code.
The code, however, did not compile any
more on any compiler on the IBM mainframe.

Finally, by summer 1997 and having solved all obstacles, 
the JADE REDUC data plus some special data selections 
were successfully copied to 600 IBM3490 cartridges.
The REFORM raw data were dropped and no longer included in the preservation process.
The 3490 cartridges were 6 times smaller than the old IBM tapes, c.f. Fig.~\ref{fig:tapes},
and carried 800 MB, i.e. 5 times more data each.
A second copy was written to 200 \oq Exabyte" cartridges (from Exabyte Corporation, a US-American 
provider of innovative storage solutions), carrying up to 2.5 GB each.
These were privately stored in the office of a caring member of JADE.
The volume of the size of a drawer was (almost) reached.

In December 2005, the Exabyte cartridge collection travelled to the Max-Planck-Institute of Physics
in Munich, conveniently packed in a flight-cabin trolley, copied to the 
Max-Planck Computing and Data Facility (MPCDF) as a very small part of the huge data space 
of the ATLAS experiment at the Large Hadron Collider at CERN, and - due to their
\oq small" size of less than 1~TB - to other discs and cloud services at the MPCDF. 

Today, at the time of writing this review,
the data fit on a TB USB 3.0 Flash Drive Memory Stick which is available  for
less than 20 Euro, reducing the drawer-size to the size of a finger tip.

In parallel, since the end of the 1980s, the data of calibrated and fully reconstructed
multihadron final
states, $\epem \rightarrow$ hadrons,  were
maintained using a private, compact 4-vector format, ZE4V \cite{ze4v}.
In 1996,  the ZE4V files were read
on the DESY IBM mainframe, converted to ASCII files, and were then 
stored at the computer centres at RWTH Aachen and - later - at the MPCDF in Garching. 

So the original JADE data were preserved. 
All of the relevant data? Apparently, a small number of files, the JADE luminosity files,
escaped preservation and apparently had been lost in the late 1990s. 
These files contained the reconstructed values of the integrated luminosities of each of 
the data taking runs, therefore being essential for calculating physical cross sections of processes 
under study. 
A worldwide search within the JADE collaboration returned no results, until Jan Olsson
found a printed version of the luminosity files at DESY.
The printout, however, was on green recycling paper and too faint for scanning and
optical character recognition, OCR.
Instead, the numbers had to be typed in a tedious effort into a text file.
Only 5 typing errors were found and corrected by a checksum routine (that was preserved in
the process of software revival, see below), and so even the luminosity files were successfully
recovered\footnote{Jan Olsson's moral on this story \cite{olsson-2009}: 
\textit{Always keep a printout as backup! (And never
throw papers away)}.}.

\subsection{Software Revival}

Generation of new detector level Monte-Carlo (MC) generated data 
sets required the revival of the entire JADE reconstruction and production software. 
The JADE software source libraries were available, with very
few exceptions\footnote{The software package for simulating the JADE muon detector  
escaped long-term archiving and today must be considered to be
inevitably lost.},
on the DESY IBM main frame, as private
copies since 1991.
In 1997, the IBM main frame was phased out and the transition to 
UNIX platforms was made. 
The original software, consisting of FORTRAN-IV, 
but also in part of SHELTRAN and MORTRAN routines, required partial 
rewriting or conversion to FORTRAN-77; 
the functionality of parts of the DESYLIB, originally written in 
assembler code, needed to be emulated with FORTRAN routines. 
\begin{figure}[htb]
  \includegraphics[width=\textwidth]{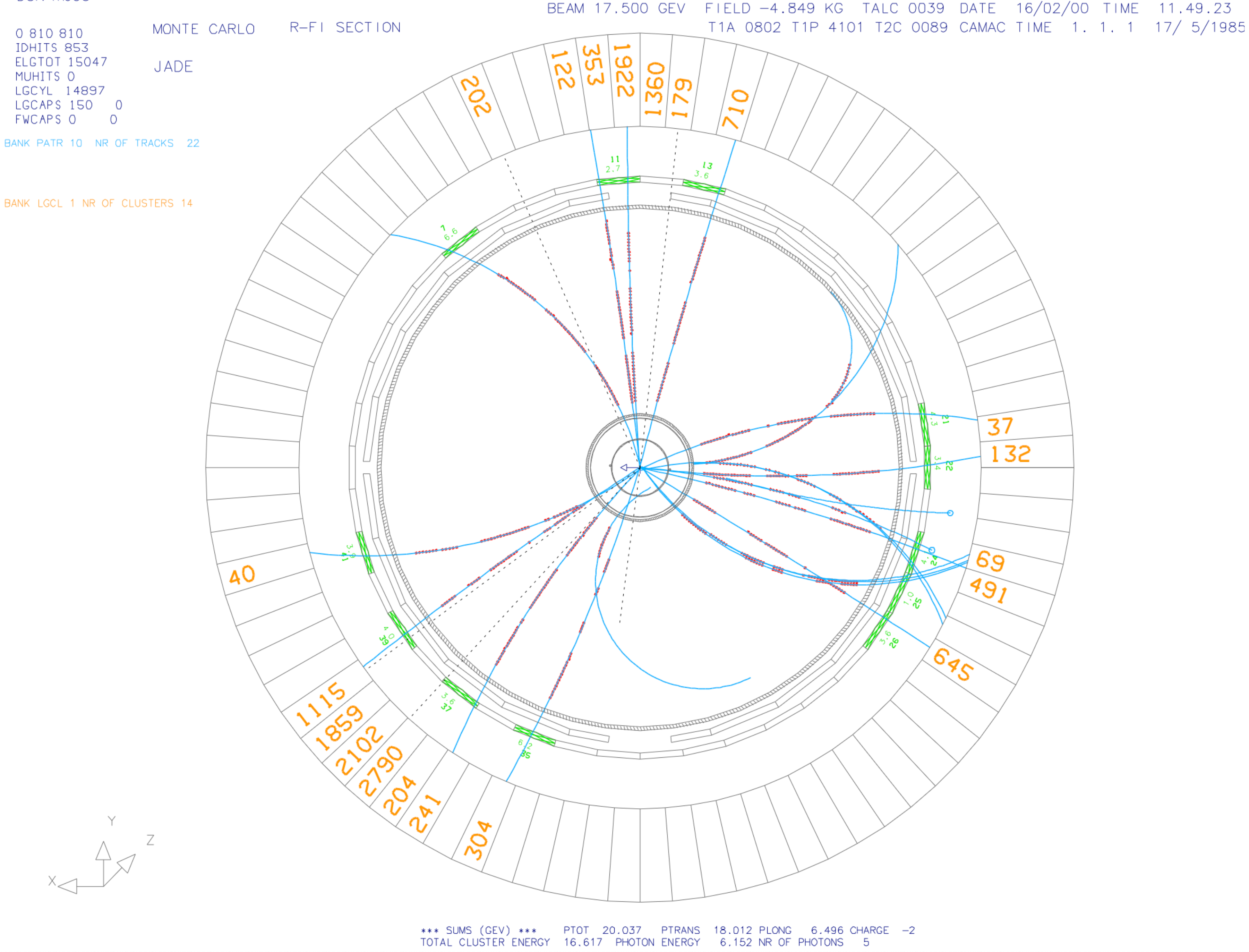}
\caption{Monte Carlo generated 3-jet event, reconstructed and displayed using the full 
revived JADE offline software chain. Note that the display is in colour, a feature that was not
available at JADE running times.}
\label{fig:MC-3jet}       
\end{figure}

The reactivation of the JADE software was completed in 1999 \cite{fernandez},
and was used for various new analysis projects at RWTH Aachen and
at MPI of Physics in Munich, until 2013. 
The new code versions were validated using various detector-level 
distributions from original and newly generated MC event samples.
An important tool of validation was the graphical display of JADE events,
which was revived to full functionality by emulating the originall
graphics and plotting software PLOT10. Since its revival, the
event display is capable of using colours, a feature that was not yet 
available at running time of the experiment, see Fig.~\ref{fig:MC-3jet}
for a coloured display of a MC-generated, fully reconstructed hadronic event.

In this first phase of software resurrection and data re-use,
the revived JADE software exclusively ran on IBM AIX machines, relying on
the fact that these systems utilise the same byte order as the IBM 370 did\footnote{The IBM370 
as well as IBM AIX machines
operated according to the \oq big-endian" convention, i.e. the highest value byte of 
a (2- or 4-byte) word
had the smallest storage address within the word.}.
It was used to generate new MC-generated data samples, based on modern 
MC generators, for correction of detector response and resolution,
for unfolding data distributions and subsequent comparison with new theoretical calculations.
The newly generated MC data sets were converted to the ZE4V format, and physics analyses
were then performed using the ZE4V files of both real and MC data.

After 2010, the restriction that JADE data analyses could only be performed on IBM AIX platforms
became increasingly inconvenient, due to the restricted availability of AIX systems and the
general move of scientific computing to LINUX systems and modern build tools.
The second, and so far final, phase of migrating the JADE software to make it compatible with
modern compilers, operating systems and build tools was and still is being executed and maintained 
by a small \oq data preservation" group at the Max-Planck-Institute of Physics 
in Munich, with Andrii Verbytskyi as the main active scientist and computing expert of the group.

This phase of data preservation so far
concentrated on using GNU Fortran and Intel Fortran, both supporting 
I/O with multiple endianness -
an essential feature to ensure functionality of the JADE software.
In order to avoid dependences on meanwhile unsupported software libraries like
CERNLIB \cite{cernlib}
and the HIGZ graphic package \cite{higz}, the required functions were emulated 
with the help of ROOT \cite{root}. 
However, the differences in the treatment of graphics in HIGZ and ROOT 
resulted in instabilities of the detector display program. 
Therefore, it is expected that JADE software will benefit from ongoing efforts 
to preserve and re-consolidate CERNLIB.

Due to the portability of the \texttt{cmake} build system \cite{cmake}, it is now possible to compile the
JADE software not only on LINUX systems but also, for the first time, on MacOSX.
The codebase was put in a public GitHub \cite{github} account, which allowed for regular 
automated builds of the JADE software on these platforms.
In order to make modern and generally very complex Monte Carlo generators compatible with
the full JADE detector simulation, a utility for converting 
HepMC \cite{hepmc} events into JADE-readable format was created,
thus avoiding the need of multiple different interfaces to MC event generators. 

\subsection{New Results from Resurrected data}
\label{sec:newresults}

With the resurrection and revival
of data and software, the second phase of JADE physics analyses started
in 1997 at RWTH Aachen, moving to MPP Munich in 2000.
It was based on analysing calibrated and fully reconstructed  
$\epem \rightarrow$ hadron events, 
preserved in the compact ZE4V 4-vector format, and on newly generated Monte-Carlo event
samples, using modern QCD shower models, processed through the full JADE detector 
simulation, reconstruction and ZE4V conversion chain.
The typical number of data events used in those studies is given in Table~\ref{tab:newdata}.

\begin{table}[htb]
\centering
\caption{The average centre-of-mass energy $\sqrt{s}$,
the energy range, data taking period, collected integrated luminosity L
and the number of selected data events,
after applying quality selection cuts, as typically used 
in the second, post-resurrection phase of JADE data analyses.}
\label{tab:newdata}      
\begin{tabular}{| c | c | r | c | r |}
\hline
$\sqrt{s}$ & energy & year & L & selected \\
(GeV) & range (GeV) &  & (pb$^{-1}$) & events \\
\hline
14.0 & 13.0-15.0 & 1981 & 1.46 & 1783 \\
22.0 & 21.0-23.0 & 1981& 2.41 & 1403 \\
34.6 & 33.8-36.0 & 1981-1982 & 61.7 & 14313 \\
35.0 & 34.0-36.0 & 1986 & 92.3 & 20876 \\
38.3 & 37.3-39.3 & 1985 & 8.28 & 1585 \\
43.8 & 43.4-46.4 & 1984-1985 & 18.8 & 4374 \\
\hline
\end{tabular}
\end{table}

Re-analyses of the JADE data were motivated by the significant increase of knowledge, 
both in theory and experimental techniques, obtained in the 1990s along with the operation of LEP,
the $\epem$ collider at CERN that operated from 1989 to 2000 at c.m. energies from 90 to 214~GeV.
Specifically, the main interest focussed on extending advanced studies of hadronic final states
and tests of QCD, see e.g. \cite{sb-physrep} for a review of QCD results from LEP,
to the lower energy data from PETRA, namely on 
\begin{itemize}
\item
measurements of new and improved observables,
like hadronic event shapes, and application of new and improved jet algorithms, 
\item
scrutiny and application of significantly improved QCD shower Monte Carlo generators, 
\item
application of advanced QCD calculations, in next-next-to-leading order (NNLO)
perturbation theory and (next-to-) leading logarithmic approximation (NLLA),
\item
precision determinations of the strong coupling parameter $\as$, and on
\item
general tests of QCD.
\end{itemize}

Until 2013, three PhD theses, ten journal publications and several 
scientific presentations at international conferences and proceedings 
articles emerged from this second phase of JADE data analyses.
The new results
were novel or superior to, but always compatible with those that had been previously published
by JADE.
In the following, two areas of major new insights from these new studies will be reviewed.

\begin{figure}[htb]
 \begin{center}
  \includegraphics[width=0.5\textwidth]{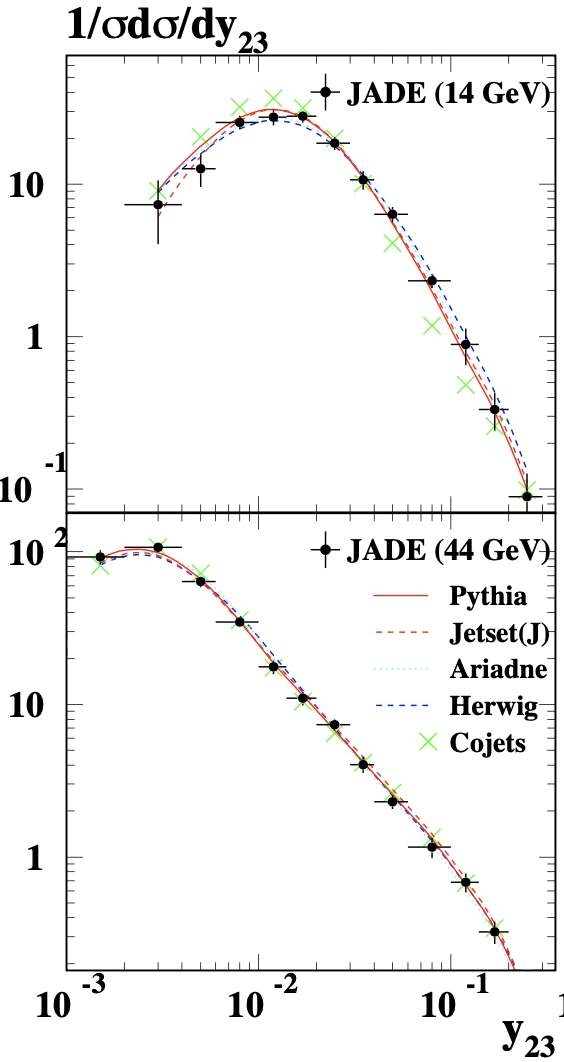}
 \end{center}
\caption{Measured distributions of $y_{23}$, the value of the jet resolution  
where the event changes from a 2-jet to a 3-jet configuration, at the lowest and highest
PETRA energies of $\sqrt{s}$ = 14 GeV and 44 GeV, compared with various QCD model
calculations using hadronisation parameters obtained from LEP data at $\sqrt{s}$ = 91.2 GeV
\cite{fernandez-thesis,kluth2003}.}
\label{fig:shapes}       
\end{figure}
\subsubsection{New Observables and QCD models: Universal Description of Hadronisation}
The first results from the revival phase of JADE were published in 1998, titled
\textit{A Study of event shapes and determinations of 
$\as$ using data of $\epem$ annihilations at $\sqrt{s}$ = 22~GeV to 44~GeV} \cite{jade-as1}.
It was based on the PhD thesis of P.A. Movilla-Fernandez \cite{fernandez-thesis}, as the first application
and use of the resurrected JADE data and revived software \cite{fernandez}.
These first results presented measured distributions of new event shape observables and jet
resolution parameters, their comparison with various improved QCD 
shower models and event generators, and the extraction of values of $\as (Q)$ at energy scales
of $Q \equiv \sqrt{s} =$ 14, 22, 34 and 44 GeV, based on QCD in next-to-leading order (NLO)
perturbation theory plus resummation of next-to-leading logs (NLLA).

None of the new observables, the improved QCD generators, the $\as$ results at the smallest PETRA
energies, nor the resummed QCD calculations had been available at PETRA
running time before.
Therefore these studies provided a wealth of new and unique 
results. 
While measurements of $\as$ will be further discussed in the next subsection, another aspect of
these results shall be emphasised here:

At PETRA running time, the low energy data at 14 and at 22 GeV were hardly used, 
mainly because MC models - at that time, based on LO or NLO QCD
only - did not provide satisfactory descriptions of the data, even if the hadronisation parameters of
those models were re-tuned at each energy.
In contrast, the new JADE studies used the new and improved QCD shower models, which
include gluon emission processes to much higher orders, down to much lower invariant masses of
$\cal{O}$(1 GeV), leaving a much smaller phase space for the nonperturbative
hadronisation process. 
Using those shower models with hadronisation parameters optimised to describe the  
LEP data around the $Z^0$ pole, leaving all parameters including the QCD scale parameter 
$\Lambda$ constant\footnote{Note that QCD predicts the running and energy dependence of 
$\as \sim \log{(\Lambda / Q)}$, with $\Lambda$ being a constant scale parameter.},
surprisingly resulted in an excellent description of JADE data 
at all PETRA energies, see Fig.~\ref{fig:shapes}. 
This observation is compatible with the QCD expectation of
a running coupling and an energy-independent description of the hadronisation process.

\begin{figure}[htb]
 \begin{center}
  \includegraphics[width=0.8\textwidth]{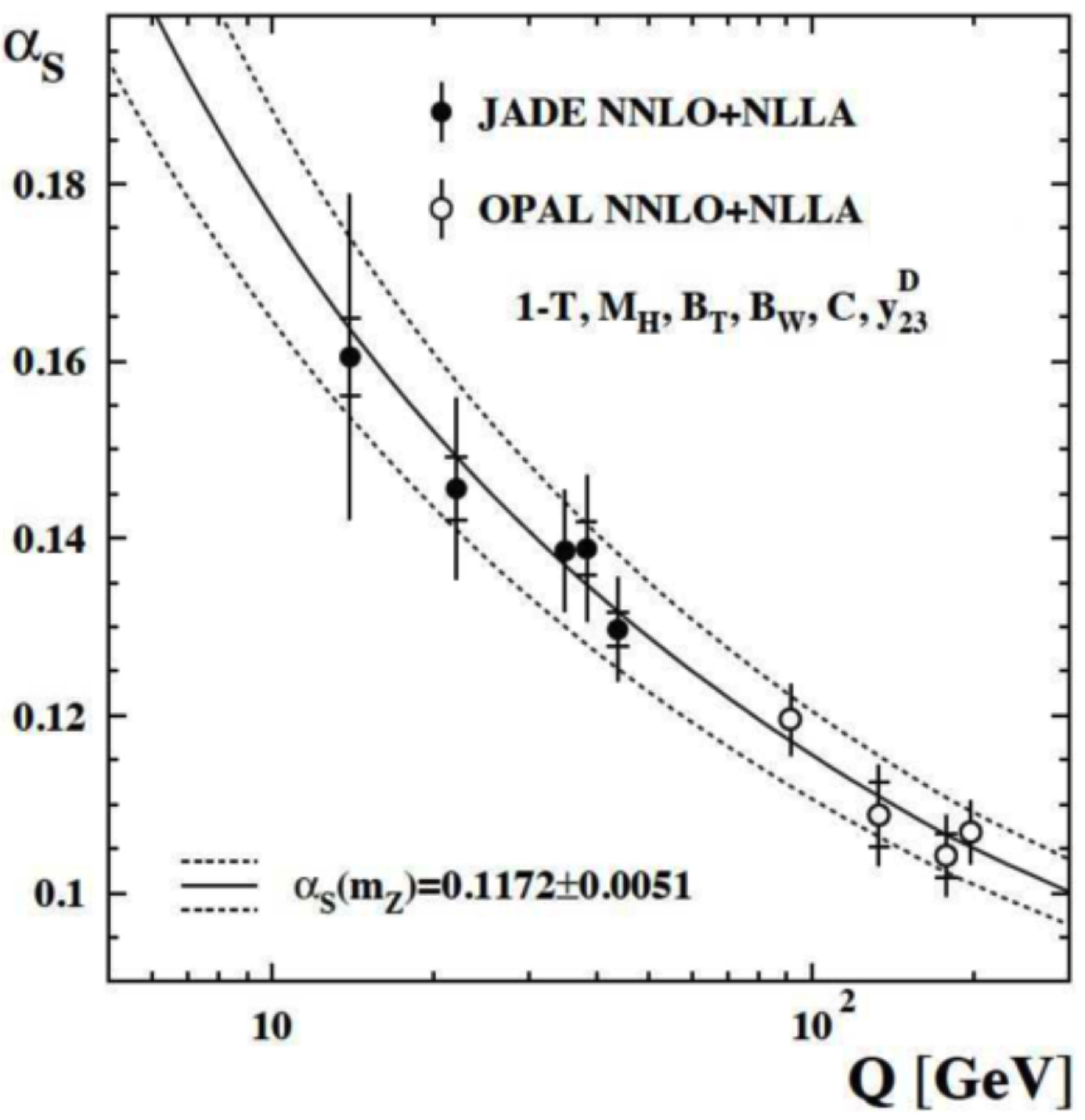}
 \end{center}
\caption{The values for $\as$ at the JADE energy points. 
The inner error bars correspond to the combined statistical and experimental errors 
and the outer error bars show the total errors. The results from $\sqrt{s}$
= 34.6 and 35 GeV have been combined for clarity. 
The full and dashed lines indicate the QCD prediction of the running coupling based on
the combined JADE NNLO result. 
The results from an NNLO analysis of OPAL data are shown as well
\cite{jade-as-nnlo,opal-alphas}.}
\label{fig:opal-alphas}       
\end{figure}
\subsubsection{Measurements of $\alpha_s$ and Signature of Asymptotic Freedom}
Most of the JADE revival studies explicitly or indirectly dealt with determinations 
of the strong coupling $\as$, owing to a special feature of the JADE data: the wide range of 
\oq low" c.m. energies where the QCD prediction of the running coupling is particularly pronounced.
Starting with the first publication mentioned in the previous subsection
\cite{jade-as1}, the next step was a precise determination from jet-production rates alone, still
in resummed NLO QCD.
This was a concerted study of both the JADE collaboration and the  
OPAL collaboration at LEP, leading to a precise, combined result of 
$\amz = 0.1187^{+0.0034}_{-0.0019}$ \cite{as-jade-opal}.

The so far \oq ultimate" determination of $\as$ from JADE data was based on a study of event shape
and jet resolution distributions and the application of the latest, state-of-the-art  QCD
predictions in complete and resummed NNLO + NLLA perturbation theory \cite{jade-as-nnlo}.
The combined results from six different event shape observables at 
six JADE centre-of-mass energies from 14 to 44 GeV, assuming the QCD prediction  
of a running coupling and converting all results the reference energy scale of the $\z0$ rest mass,
is $\amz = 0.1172\pm 0.0006(stat.) \pm 0.0020(exp.) \pm 0.0035(had.) \pm 0.0030(theo.)$.
This result significantly contributes to and is compatible with the current world average value \cite{pdg2021}.

More important, the values obtained for the different c.m. energy scales are 
compatible with the predicted QCD running of the coupling, and with the results obtained 
from an identical (but separate) analysis of the OPAL data at LEP \cite{opal-alphas}, 
see Fig.~\ref{fig:opal-alphas}.
In fact, the JADE results alone exclude the $absence$ of running with a significance and confidence level
of 99\% - a wonderful confirmation and significant strengthening 
of JADE's early \textit{Investigation of the Energy Dependence
of the Strong Coupling Strength} \cite{Bethke1988}, see Sec.~\ref{sec:jets}.

%% file: Sec-05.tex
\section{Summary and closing remarks}
\label{sec:5}

\begin{figure}[htb]
 \begin{center}
  \includegraphics[width=1.0\textwidth]{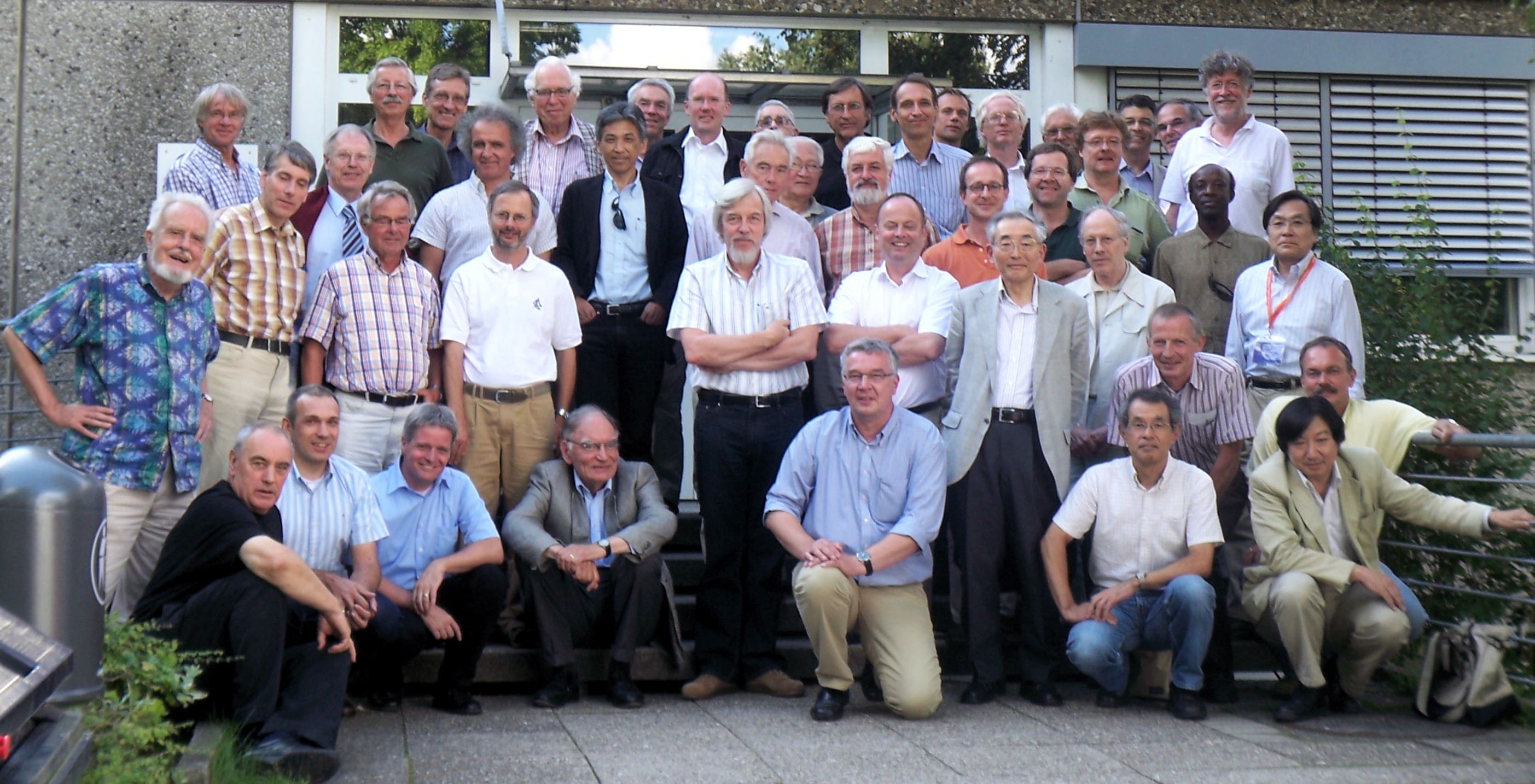}
 \end{center}
\caption{The JADE Collaboration at its reunion in August 2009.
From left to right:
-front row, sitting/bending:
Robin Marshall, Klaus Kleinworth, G\"unter Eckerlin, Rolf Felst, 
Siggi Bethke, Tomio Kobayashi, Alfred Petersen, Sachio Komamiya, Manfred Zimmer
- first row standing:
Hanns Krehbiel$^\dagger$, Peter Warming, Wulfrin Bartel$^\dagger$, Uwe Schneekloth, Rolf Heuer,
Karlheinz Meier$^\dagger$, Sakue Yamada, Dieter Haidt, Hiroshi Takeda
- second row standing:
Hans von der Schmitt, Jan Olsson, Farid Ould-Saada, Tatsuo Kawamoto, 
Hans Rieseberg, Hajime Matsumura, Karl Ambrus, Michael Kuhlen, Austin Ball, 
Andreas Dieckmann, Harrison Prosper,
- back row(s) standing:
J\"urgen von Krogh, Eckhard Elsen, Joachim Heintze$^\dagger$, Rainer Ramcke$^\dagger$, Norbert Magnussen,
Paul Murphy$^\dagger$, G\"otz Heinzelmann, Henning Kado, Stefan Kluth, Roger Barlow, unidentified person,
Hugh McCann, Herbert Drumm, Albrecht Wagner.
}
\label{fig:collab}       
\end{figure}

From today's perspective, JADE was a rather small collaboration, consisting - in total and summed over
the active lifetime of the experiment - of about 130 scientists, technicians and engineers from 
8 institutions in Germany, England, Japan and the USA \cite{jade-all}. 
The detector, though, already exhibited key features of today's large experiments in
high energy physics, concerning hermeticity and sensitivity to detect and measure 
the features of charged leptons, photons, charged particles and hadron jets.
With only about 50 to 70 collaborators at a given time, the fast planning and construction, 
the efficient operation, detector maintenance and upgrade, and data analysis was 
challenging but successfully accomplished.
The close and personal collaboration between all members 
was a unique experience, an enormous pleasure and privilege,
and door-opener for many future careers.
\textit{JADE is us and we are JADE} is the motto guiding the relation
and the still ongoing contact between
its members.
Fig.~\ref{fig:collab} shows the members of the JADE collaboration at their reunion at DESY in 
August 2009.

Together with its friendly competitor experiments CELLO, MARK-J, PLUTO and TASSO,
JADE co-discovered the gluon, tested the electro-weak standard model of
particle physics, explored many aspects of hadron, lepton and photon production at PETRA,
established significant tests of QCD and measurements of the QCD coupling,
advanced two-photon physics, performed searches for Supersymmetry and other signals of
physics beyond the standard model.

JADE's pioneering and  unique contributions were 
the development and establishment of the JADE jet finding algorithm that was widely 
used and further developed thereafter, the evidence for string fragmentation, and the
evidence for asymptotic freedom of quarks and gluons.
The results on the energy dependence of $\as$ and the signature of asymptotic freedom 
were a decisive element of granting the 2004 Nobel Prize in Physics \cite{nobel2004} to the
fathers of Asymptotic Freedom of Quarks and Gluons, D.J. Gross, F. Wilczek and H.D. Politzer.

The long-term preservation of JADE data and the revival of the JADE software, the second phase of data
analysis starting more than 10 years after shut-down of the experiment, and 
last not least the decision to save and convert JADE's heritage to "open data", for 
future unrestricted and public use, are unique key features of JADE. 
They raised international 
recognition and serve as motivation for large scale efforts of data preservation
in particle physics \cite{dphep} and beyond.

The question \textit{was it worth the effort?} can be answered with a clear \textit{yes}!
Besides being a show-case for similar initiatives currently going on, the scientific output of
re-analysing old data with advanced methods and knowledge is more than worth the effort,
and will appear in the future, for many cases again.
Such initiatives, however, deserve and justify more concerted effort and resources than 
was available for the case of JADE. 
For current and future scientific endeavours that generate orders of magnitude larger amounts of data and
software, and that may operate for significantly longer time durations,
it will be vital to organise, prepare and profit from data and software preservation already during
active life-time of the project, see e.g. \cite{atlas-dphep}.

In the case of JADE, retroactive data and software preservation was only
possible through personal initiatives of a few motivated and passionate individuals.
Jan Olsson, one of the key persons in this process, commented this fact 
\cite{olsson-2009}
by citing Erich K\"astner,
a famous German writer, publicist and cabaret artist:

\begin{center}
\textit{Es gibt nichts Gutes, \\
  ausser man tut es! \footnote{engl.: \textit{There is nothing good unless you do it.}}
  }
  \end{center}
  
May JADE and the awareness for data preservation live forever!

\vskip0.5cm
\noindent
\textbf{Acknowledgements}
We are deeply grateful to our JADE colleagues for the extremely successful and pleasant
cooperation, and for their open team spirit and continued sense of belonging since the close-down
of the experiment. 
We especially thank Jan Olsson for his tireless dedication to preserving the heritage of JADE,
and Pedro Movilla-Fernandez and Andrii Verbytskyi for their invaluable contributions to reactivate 
and ensure future use of the JADE software.
We thank E. Elsen, R. Felst, D. Haidt, R. Marshall, J. Olsson, A. Verbytskyi and S. Yamada for their corrections, suggestions, additional information and memorable stories.
SB acknowledges financial support of the data preservation project through the
Deutsche Forschungsgemeinschaft (DFG) and
the Max-Planck-Society (MPG).